\documentclass[twoside]{article}
\usepackage{qic,epsfig,amsfonts,amsmath,amssymb}
\usepackage[top=30truemm,bottom=30truemm,left=20truemm,right=20truemm]{geometry}

\newtheorem{theo}{Theorem}

\newtheorem{prop}{Proposition}
\begin{document}

\normalsize\textlineskip
\thispagestyle{empty}
\setcounter{page}{1}

\vspace*{0.88truein}
\alphfootnote

\fpage{1}

\centerline{\bf
The time-averaged limit measure of the Wojcik model}
\vspace*{0.37truein}\centerline{\footnotesize
TAKAKO ENDO}
\vspace*{0.015truein}
\centerline{\footnotesize\it Department of Physics,}
\baselineskip=10pt
\centerline{\footnotesize\it Ochanomizu University,  2-1-1 Ohtsuka, Bunkyo, Tokyo, 112-0012, Japan}
\vspace*{10pt}
\centerline{\footnotesize
NORIO KONNO}
\vspace*{0.015truein}
\centerline{\footnotesize\it Department of Applied Mathematics, Faculty of Engineering,}
\baselineskip=10pt
\centerline{\footnotesize\it Yokohama National University, Hodogaya, Yokohama, 240-8501, Japan}
\vspace*{10pt}

\vspace*{0.225truein}
\vspace*{0.21truein}
\begin{abstract}
We investigate ``the Wojcik model" introduced and studied by Wojcik et al. \cite{wojcik}, which is a one-defect quantum walk (QW) 
having a single phase at the origin. 
They reported that giving a phase at one point causes an astonishing effect for localization.
There are three types of measures having important roles in the study of QWs: 
time-averaged limit measure, weak limit measure, and stationary measure.   
The first two measures imply
 a coexistence of localized behavior and the ballistic spreading
in the QW. 
As Konno et al. \cite{segawa} suggested, the time-averaged limit and stationary measures are closely related to 
each other for some models. 
In this paper, we focus on a relation between the two measures for the Wojcik model. 
The stationary measure was already obtained by our previous work \cite{watanabe}.
Here, we get the time-averaged limit measure by several methods. 
Our results show that the stationary measure is a special case of the time-averaged limit measure.

\end{abstract}
\vspace*{10pt}
\vspace*{3pt}
\vspace*{1pt}\textlineskip    
\section{Introduction}
\label{intro}

As a quantum counterpart of the random walk, quantum walks (QWs) describe many kinds of phenomena in quantum scale \cite{kitagawa,ahlbrecht}.
There are two distinct 
types of QWs, one is the discrete time walk and the other is the continuous one.
Discrete time QWs have been intensively studied in \cite{konno5,venegas}. 
Here, we focus on a two-state discrete time QW in one dimension. 
The two-state corresponds to left and right chiralities, respectively \cite{konno1}.
It has been reported that one-dimensional discrete time QWs have characteristic properties, that is, localization 
and the ballistic spreading. There are two kinds of limit theorems to show the asymptotic behavior of the QWs: 
the time-averaged limit theorem corresponding to localization, and the weak limit theorem corresponding to the ballistic spreading.
In this paper, we say that the walk starting from the origin exhibits localization if and only if its time-averaged limit measure at the origin is strictly positive.
As Konno et al. \cite{segawa} reported, the time-averaged limit and stationary measures are closely related to
 each other. 
Therefore, we clarify the relation between the two measures for a suitable QW model.
Wojcik et al. \cite{wojcik} showed that giving a phase at a single point in the QW on the line exhibits an astonishing
 localization effect.
In this paper, we call the model ``the Wojcik model".
Our previous work \cite{watanabe} gave a stationary measure of the model, and 
this paper is a sequential work of \cite{watanabe}.
We present the time-averaged limit measure, derived from the pass counting method \cite{konno2,konno4}, the CGMV method \cite{cantero}, and the generating 
function method \cite{segawa} explained in Sect. $5$. Our result implies that 
the stationary measure is a special case of the time-averaged limit measure. 

The rest of this paper is organized as follows. 
Section $2$ gives the definition of the time-averaged limit measure and 
localization for the discrete time QW starting at the origin. 
In Sect. $3$, we introduce the Wojcik model and present our main results, Theorems $1$ and $2$. 
Section $4$ is devoted to the result based on the CGMV method.
We give the proofs of Lemma $1$ in Sect. $5$ and Theorem $2$ in Sect. $6$, respectively. 
Appendix A gives the proof of Theorem $1$, and Appendix B presents the proof of Lemma $7$.

\section{The time-averaged limit measure and localization}
In this section, we introduce the time-averaged limit measure and define localization for the QW starting 
at the origin.
First, we give the notation of the space-inhomogeneous QWs on the line. 
The walker has a coin state described by a two-dimensional vector which is called ``the probability amplitude".
We define the coin state at position $x$ and time $n$ by
\[\Psi_{n}(x)=
\begin{bmatrix}
\Psi^{L}_{n}(x)\\
\Psi^{R}_{n}(x)
\end{bmatrix}.\]
The upper and lower elements express left and right chiralities, respectively.
Let
\[\Psi_{n}= {}^T\![\ldots,\Psi_{n}^{L}(-1),\Psi_{n}^{R}(-1),\Psi_{n}^{L}(0),\Psi_{n}^{R}(0),\Psi_{n}^{L}(1),\Psi_{n}^{R}(1),\ldots ],\]
where $T$ means the transposed operation. 
The time evolution is defined by its initial coin state $\Psi_{0}$ and $2\times 2$ unitary matrices $U_{x}\;(x\in\mathbb{Z})$ :

\[U_{x}=\begin{bmatrix}
a_{x}&b_{x}\\
c_{x}&d_{x}\\
\end{bmatrix},\]\\
where subscript $x\in\mathbb{Z}$ denotes the location.
Then the evolution is determined by the following recurrence formula:
\[\Psi_{n+1}(x)= P_{x+1}\Psi_{n}(x+1)+ Q_{x-1}\Psi_{n}(x-1),\]
where
\[ 
P_{x}=\begin{bmatrix} 
a_{x} & b_{x} \\
0 & 0 
\end{bmatrix},\quad
Q_{x}=\begin{bmatrix}
0 & 0 \\
c_{x} & d_{x}
\end{bmatrix}.\]
Note that $P_{x}$ (resp. $Q_{x}$) expresses that the walker moves to the left (resp. right) at position $x$ 
in each time step. 
Let $\mathbb{R}_{+}=[0,\infty)$. 
Then for 
\[\Psi_{n}= {}^T\!\left[\ldots,\begin{bmatrix}
\Psi_{n}^{L}(-1)\\
\Psi_{n}^{R}(-1)\end{bmatrix},\begin{bmatrix}
\Psi_{n}^{L}(0)\\
\Psi_{n}^{R}(0)\end{bmatrix},\begin{bmatrix}
\Psi_{n}^{L}(1)\\
\Psi_{n}^{R}(1)\end{bmatrix},\ldots\right]\in(\mathbb{C}^{2})^{\mathbb{Z}},\]
we define a map $\mu_{n}:\mathbb{Z}\to[0,\infty]$ as
\[\mu_{n}(x)=|\Psi_{n}^{L}(x)|^{2} + |\Psi_{n}^{R}(x)|^{2}
\quad(x\in\mathbb{Z}).\]
Our interest in this paper is the sequence of measures:
\[\{\mu_{0},\mu_{1},\mu_{2},\ldots\}.\]
If $\mu_{n}$ is a probability measure,
let $X_{n}$ be a random variable defined by $\mu_{n}$, that is, for $x\in\mathbb{Z}$, 
\[P(X_{n}=x)=\mu_{n}(x).\]
Now we introduce the time average of $\mu_{n}(x)$ and its limit. 
The time average of $\mu_{n}(x)$ is defined by
\[\overline{\mu}_{T}(x)=\frac{1}{T}\sum_{n=0}^{T-1}\mu_{n}(x),\]
and if the limit exists, we define the limit of $\overline{\mu}_{T}(x)$ by
\begin{align}\overline{\mu}_{\infty}(x)=\lim_{T\to\infty}\overline{\mu}_{T}(x)
=\lim_{T\to \infty}\frac{1}{T}
\sum^{T-1}_{n=0}P(X_{n}=x).\label{(2)}\end{align}
Here we put 
\begin{align}
\overline{{\mathcal M}}_{\infty}=\{\overline{\mu}_{\infty}=\overline{\mu}_{\infty}^{\Psi_{0}}
\in\mathbb{
Z}_{+}^{\mathbb{Z}}\setminus\{{\bf 0}\}:\Psi_{0}\in\mathbb{C}^{\mathbb{Z}}\},\label{(3)}
\end{align}
where $\overline{\mu}_{\infty}^{\Psi_{0}}$ represents the dependence on the initial state $\Psi_{0}$ and 
$\{{\bf 0}\}={}^T\![\ldots,0,0,0,\ldots]$. 
We call the element of $\overline{{\mathcal M}}_{\infty}$ the time-averaged limit measure of the QW.
Then, localization for discrete time QW is defined as follows.

\par\indent
\begin{definition}We say that
localization for the QW starting at the origin happens if
\[\overline{\mu}_{\infty}(0)>0.\]
\end{definition}

\section{Model and main results}
\subsection{Model}
In this paper, we treat a space-inhomogeneous QW, ``the Wojcik model", introduced by Wojcik et al. \cite{wojcik}, whose time evolution is defined by the unitary matrices 
$U_{x}\;(x\in\mathbb{Z})$ as follows.
\begin{align}U_{x}=\left\{\begin{array}{ll}
H&(x\in\mathbb{Z}\setminus\{0\}),\\
\omega H&(x=0),\end{array}\right.\label{(4)}\end{align}
where $\omega=e^{2\pi i\phi}$ with $\phi\in(0,1)$.
The model has a weight $e^{2\pi i\phi}$ at the origin. 
Here,  $H$ is ``the Hadamard matrix":
\begin{align}
H=\dfrac{1}{\sqrt{2}}\begin{bmatrix} 1&1\\ 1& -1\end{bmatrix}.\label{(5)}
\end{align}
In particular, if $\phi\to 0$, then the Wojcik model becomes space-homogeneous 
and is equivalent to the well-known Hadamard walk which is one of the most intensively studied QWs. 
We should note that Konno et al. \cite{segawa} treated the QW in which $\det(U_{x})$ does not depend on the position $x\in\mathbb{Z}$.
However, the Wojcik model has
\[\det(U_{0})=-1,\quad\det(U_{x})=-\omega^{2}\;(\neq-1\;if\; x\in\mathbb{Z}\setminus\{0\}).\]
In this paper, we assume that the walk starts at the origin with the initial coin state 
$\varphi={}^T\![\alpha,\beta]$, where $\alpha,\beta\in\mathbb{C}$ and $|\alpha|^{2}+|\beta|^{2}=1.$

\subsection{Main result ${\bf 1}$: Time-averaged limit measure at the origin}
In this subsection, we give the time-averaged limit measure at the origin. 
We should remark that we treat the measure for $|x|\geq1$ case in subsection $3.3$.  
Let us consider the initial coin states $\varphi={}^T\![1/\sqrt{2},i/\sqrt{2}]$, 
or $\varphi={}^T\![1/\sqrt{2},-i/\sqrt{2}]$ for a while.
At first, we focus on the Hadamard walk ($\phi\to0$ case).
For the initial coin state, the probability distribution of the walk is symmetric for the origin at 
any time.
If $\mu_{n}$ is a probability measure, let $X_{n}$ be the random variable of the walk for the position $x$ at time $n$. We compute the return probability at 
time $n$, which we denote it as $r_{n}^{(H)}(0)=P(X_{n}=0)$. We should note that $r_{2n+1}^{(H)}(0)=0\;(n\geq0)$.
By a brief calculation, we have
\begin{eqnarray*}
r^{(H)}_{2}(0)=0.5,\quad r^{(H)}_{4}(0)=0.125,\quad r^{(H)}_{6}(0)=0.125,\quad
r^{(H)}_{8}(0)=0.07031,\nonumber\\ r^{(H)}_{10}(0)=0.07031,\quad
r^{(H)}_{12}(0)=0.04882,\quad r^{(H)}_{14}(0)=0.04882,\ldots.\label{(35)}
\end{eqnarray*}
In fact, we see
\begin{align}\lim_{n\to\infty}r^{(H)}_{2n}(0)=0,\label{37}\end{align}
for example, see \cite{konno2}.
Equation (\ref{37}) suggests that the Hadamard walk ($\phi\to0$ case) does not show localization .
From now on, we consider a space-inhomogeneous case, that is, $\phi\in(0,1)$ case.
By a simple calculation, we have the same probability measure as that of the Hadamard walk at time $n=1,2,3$ for the initial coin states $\varphi={}^T\![1/\sqrt{2},\eta i/\sqrt{2}]\;(\eta=1,-1)$. However, 
we see that the probability measure at time $n=4$ depends on the parameter $\phi$. 
Actually, we have
\begin{eqnarray*}
P(X_{4}=-4)=P(X_{4}=4)\!\!\!&=&\!\!\!\dfrac{1}{16},\quad
P(X_{4}=-2)=P(X_{4}=2)=\dfrac{2(2+E)}{16},\nonumber\\
P(X_{4}=0)\!\!\!&=&\!\!\!\dfrac{2(3-2E)}{16},\label{(38)}
\end{eqnarray*}
where $E=C+\eta S\;(\eta=1,-1),\;C=\cos(2\pi\phi)$, and $S=\sin(2\pi\phi)$.
Hereafter, we present the time-averaged limit measure at the origin for the Wojcik model.
Let \begin{align*}
\Psi_{2n}(0)=\left[\begin{array}{c}\Psi^{L}_{2n}(0)\\ \Psi^{R}_{2n}(0)\end{array}\right]
\end{align*} be the probability amplitude at time $2n$ at the origin.
Then, we obtain an explicit expression for $\Psi_{2n}(0)$ as follows.
\par\indent\par\noindent
\begin{lemma}
\label{wojcik-lemma1}
Let $\varphi=\varphi(\eta)={}^T\![1/\sqrt{2},\eta i/\sqrt{2}]\;(\eta=1,-1)$ be the initial coin state.
Then, we have 
\begin{align*}
\Psi_{2n}(0)=\dfrac{1}{\sqrt{2}}\sum^{n}_{k=1}\sum_{\scriptstyle (a_1, \ldots , a_k) \in (\mathbb{Z}_{>})^k : \atop 
\scriptstyle a_1 + \cdots + a_k =n}
\left(\prod^{k}_{j=1}r^{\ast}_{2a_{j}-1}\right)
\left(\dfrac{\omega(-1+\eta i)}{2}\right)^{k}
\left[\begin{array}{c}1\\ \eta i\end{array}\right],
\end{align*}
for $n\geq1$, where $\mathbb{Z}_{>}=\{1,2,\cdots\}$, and 
\begin{align*}
\sum^{\infty}_{n=1}r_{n}^{\ast}z^{n}=\dfrac{-1-z^{2}+\sqrt{1+z^{4}}}{z}.
\end{align*}
\end{lemma}
We prove Lemma \ref{wojcik-lemma1} in Sect. \ref{wojcik-lemma1-proof}.
Noting that the return probability is defined as $P(X_{2n}=0)=\|\Psi_{2n}(0)\|^{2}=|\Psi^{L}_{2n}(0)|^{2}+|\Psi^{R}_{2n}(0)|^{2}$, we have
\par\indent
\par\noindent
\begin{lemma}
\label{wojcik-lemma2}
Let $\varphi=\varphi(\eta)={}^T\![1/\sqrt{2},\eta i/\sqrt{2}]\;(\eta=1,-1)$ be the initial coin state.
Then the limit of the return probability at time $2n$ for the parameter $\phi$ is given as follows.
\begin{eqnarray*}
c(\phi)\!\!\!&=&\!\!\!\lim_{n\to\infty}r_{2n}(0)\nonumber\\
\!\!\!&=&\!\!\!4\left(\dfrac{1-\sqrt{2}C_{-}}{3-2\sqrt{2}C_{-}}\right)^{2}I_{(1/4,1)}(\phi)I_{\{1\}}(\eta)
+4\left(\dfrac{1-\sqrt{2}C_{+}}{3-2\sqrt{2}C_{+}}\right)^{2}I_{(0,3/4)}(\phi)I_{\{-1\}}(\eta),\label{(42)}
\end{eqnarray*}
where $I_A (x)=1 \>(x \in A) \>,\; I_{A}(x)=0 \> (x \not\in A)$, and
\begin{align*}
C_{-} 
&= \cos \left( 2 \pi \phi - \frac{\pi}{4} \right) = \frac{\sqrt{2}}{2} \left\{ \cos ( 2 \pi \phi) + \sin ( 2 \pi \phi) \right\},
\\
C_{+} 
&= \cos \left( 2 \pi \phi + \frac{\pi}{4} \right) = \frac{\sqrt{2}}{2} \left\{ \cos ( 2 \pi \phi) - \sin ( 2 \pi \phi) \right\}.
\end{align*}
\end{lemma}
Interestingly, when $\phi\in(0,1)$, we see that the inequality
\begin{align*}c(\phi)>0\label{(43)}\end{align*}
holds except for $\eta=1$ with $\phi\in(0,1/4]$ or $\eta=-1$ with $\phi\in[3/4,1)$. 
On the other hand, when $\phi\to0$, we have $c(\phi)=0$ which implies that the Hadamard walk does not exhibit localization. 

As for the general case, that is, for the initial coin state 
$\varphi={}^T\![\alpha,\beta]\;(\alpha,\beta\in\mathbb{C},\;|\alpha|^{2}+|\beta|^{2}=1)$, 
we can obtain the probability amplitudes and the time-averaged limit measure by Lemma \ref{wojcik-lemma1}.
Now, we should note that for the general initial coin state $\varphi$, we have
\begin{align*}
\Psi_{2n} (0) 
&= \frac{1}{\sqrt{2}} \> \sum_{k=1}^n \sum_{\scriptstyle (a_1, \ldots , a_k) \in (\mathbb{Z}_{>})^k : \atop \scriptstyle a_1 + \cdots + a_k =n} \left( \prod_{j=1}^k \Xi^{\ast}_{2 a_j} \right) \> \varphi
\\
& = \frac{1}{\sqrt{2}} \> \sum_{k=1}^n \sum_{\scriptstyle (a_1, \ldots , a_k) \in (\mathbb{Z}_{>})^k : \atop \scriptstyle a_1 + \cdots + a_k =n} \left( \prod_{j=1}^k r^{\ast}_{2 a_j-1} \right) 
\> \left( \frac{\omega}{2} \right)^k \> 
\left[
\begin{array}{cc}
-1 & 1 \\
-1 & -1
\end{array}
\right]^k \> 
\left[
\begin{array}{cc}
\alpha \\
\beta
\end{array}
\right]
\end{align*}
for $n \ge 1$, where $\mathbb{Z}_{>} = \{1,2, \ldots \}$ and
\begin{align*}
\sum_{n=1}^{\infty} \> r_{n}^{\ast} z^n = \frac{-1 - z^2 + \sqrt{1 + z^4}}{z}.
\end{align*}
Here we should note 
\begin{align*}
\left[
\begin{array}{cc}
-1 & 1 \\
-1 & -1
\end{array}
\right]^k \> 
\left[
\begin{array}{cc}
\alpha \\
\beta
\end{array}
\right]
=
\left[
\begin{array}{cc}
(-1+i)^k \left( \dfrac{\alpha - i \beta}{2} \right) + (-1-i)^k \left( \dfrac{\alpha + i \beta}{2} \right) \\
(-1+i)^k \>i \> \left( \dfrac{\alpha - i \beta}{2} \right) + (-1-i)^k \> (-i) \> \left
( \dfrac{\alpha + i \beta}{2} \right)
\end{array}
\right].
\end{align*}
Therefore, we obtain the concrete expression of $\Psi_{2n}(0)$ for the general initial coin state $\varphi$ as follows.
\par\indent
\par\noindent
\begin{lemma}
\label{wojcik-lemma3}
For the initial coin state $\varphi = {}^T [\alpha, \beta] \; (\alpha, \beta \in \mathbb{C}, |\alpha|^2 + |\beta|^2 =1),$
we have 
\begin{align*}
\Psi_{2n} (0) 
&= \left( \frac{\alpha - i \beta}{2} \right)  \sum_{k=1}^n \sum_{\scriptstyle (a_1, \ldots , a_k) \in (\mathbb{Z}_{>})^k : \atop \scriptstyle a_1 + \cdots + a_k =n} \left( \prod_{j=1}^k r^{\ast}_{2 a_j-1} \right) 
\> \left( \frac{\omega(-1+i)}{2} \right)^k\> 
\left[
\begin{array}{cc}
1\\
i
\end{array}
\right]
\\
&+ \left( \frac{\alpha + i \beta}{2} \right)\sum_{k=1}^n \sum_{\scriptstyle (a_1, \ldots , a_k) \in (\mathbb{Z}_{>})^k : \atop \scriptstyle a_1 + \cdots + a_k =n} 
\left( \prod_{j=1}^k r^{\ast}_{2 a_j-1} \right) 
\> \left( \frac{\omega(-1-i)}{2} \right)^k  \> 
\left[
\begin{array}{cc}
1 \\
-i
\end{array}
\right],
\end{align*}
for $n \ge 1$.
\end{lemma}
\par\indent
\par\noindent
Thus, multiplying the results by $(\alpha - \eta i \beta)/\sqrt{2}\;(\eta=1,-1)$ for each initial coin state
$\varphi = \varphi (\eta) = {}^T [1/\sqrt{2},\eta i/\sqrt{2}] \;
(\eta =1,-1)$ and then summing the results each other gives the 
time-averaged limit measure. By Lemma $3$, we obtain the following one of our main results.
\par\indent
\par\noindent
\begin{theo}
\label{wojcik-theo1}
\begin{enumerate}
\item
For the initial coin state $\varphi = {}^T [\alpha, \beta] \> (\alpha, \beta \in \mathbb{C}, |\alpha|^2 + |\beta|^2 =1)$, we have
\begin{align*}
&\Psi_{2n} ^{(L,\Re)} (0) \sim (\alpha - i \beta) \> \frac{1-E_{+}}{3-2E_{+}} \cos (n \theta_0) I_{(1/4,1)} (\phi) 
+ (\alpha + i \beta) \> \frac{1-E_{-}}{3-2E_{-}} \cos (n \theta_0) I_{(0,3/4)} (\phi),\\
&\\
&\Psi_{2n} ^{(L,\Im)} (0) \sim (\alpha - i \beta) \> \frac{1-E_{+}}{3-2E_{+}} \frac{S-C}{|S-C|} \sin (n \theta_0) I_{(1/4,1)} (\phi)
+ (\alpha + i \beta) \> \frac{1-E_{-}}{3-2E_{-}} \frac{S+C}{|S+C|} \sin (n \theta_0) I_{(0,3/4)} (\phi),\\
&\\
&\Psi_{2n} ^{(R,\Re)} (0) 
\sim - (\alpha - i \beta) \> \frac{1-E_{+}}{3-2E_{+}} \frac{S-C}{|S-C|} \sin (n \theta_0) I_{(1/4,1)} (\phi).
+ (\alpha + i \beta) \> \frac{1-E_{-}}{3-2E_{-}} \frac{S+C}{|S+C|} \sin (n \theta_0) I_{(0,3/4)} (\phi),\\
&\\
&\Psi_{2n} ^{(R,\Im)} (0) 
\sim (\alpha - i \beta) \> \frac{1-E_{+}}{3-2E_{+}} \cos (n \theta_0) I_{(1/4,1)} (\phi)
- (\alpha + i \beta) \> \frac{1-E_{-}}{3-2E_{-}} \cos (n \theta_0) I_{(0,3/4)} (\phi),
\end{align*}
\par\noindent
for $n \geq 1$, where
\begin{eqnarray*}
\Psi^{(j,\Re)}_{2n}(0)\!\!\!&=&\!\!\!\Re\left({\Psi^{(j)}_{2n}(0)}\right),\quad\Psi^{(j,\Im)}_{2n}(0)=\Im\left({\Psi^{(j)}_{2n}(0)}\right)\;(j=L,R),\\
&\\
E_{\pm} \!\!\!&=&\!\!\! C \pm S= \cos (2 \pi \phi) \pm  \sin (2 \pi \phi),\\
&\\
\cos\theta_{0}\!\!\!&-&\!\!\!\dfrac{2(1-E)^{2}}{3-2E},\quad\sin\theta_{0}=\dfrac{(2-E)|S-C|}{3-2E}.
\end{eqnarray*}
\noindent
Here, we should note that $\Re(z)$ is the real part and $\Im(z)$ is the imaginary part of $z\;(z\in\mathbb{C})$.
\noindent
Moreover, we have
\item
\begin{eqnarray*}
\overline{\mu}_{\infty} (0) 
\!\!\!&=&\!\!\!\lim_{n \to \infty} \> \frac{r_{2n} (0)}{2} 
\\
\!\!\!&=&\!\!\!\left( \dfrac{1- E_{+}}{3- 2E_{+}} \right)^2 \> |\alpha - i \beta|^2 \> I_{(1/4,1)} (\phi)+ \left(\dfrac{1- E_{-}}{3- 2E_{-}} \right)^2 \> |\alpha + i \beta|^2 \> 
 I_{(0,3/4)} (\phi).
\\
\!\!\!&=&\!\!\!\left( \dfrac{1- \sqrt{2}C_{-}}{3- 2\sqrt{2}C_{-}} \right)^2 \> |\alpha - i \beta|^2 \> I_{(1/4,1)} (\phi)
+ \left( \dfrac{1- \sqrt{2}C_{+}}{3- 2\sqrt{2}C_{+}} \right)^2 \> |\alpha + i \beta|^2 \>  I_{(0,3/4)} (\phi),
\end{eqnarray*}
where
\begin{eqnarray*}
C_{\pm} = \cos \left( 2 \pi \phi \pm \frac{\pi}{4} \right) = \frac{1}{\sqrt{2}} \left\{ \cos ( 2 \pi \phi) \mp \sin ( 2 \pi \phi) \right\}.
\end{eqnarray*}

\end{enumerate}
\end{theo}

The proof of Theorem $1$ appears in Appendix A.

\subsection{Main result {\bf $2$}: Time-averaged limit measure for general $x\in\mathbb{Z}$}
Let us consider the Wojcik model starting at the origin with the initial coin state $\varphi={}^T\![\alpha,\beta]$, where $\alpha,\beta\in\mathbb{C}$ with $|\alpha|^{2}+|\beta|^{2}=1$.
In this subsection, we present the time-averaged limit measure $\overline{\mu}_{\infty}(x)\;(x\in\mathbb{Z})$.

\par\indent
\begin{theo}
\label{main-result2}
\begin{enumerate}
\item $x=0$.
\[\overline{\mu}_{\infty}(0)=\overline{\mu}^{(1)}(0)+\overline{\mu}^{(2)}(0).\]
\item $x\neq0$.
\[\overline{\mu}_{\infty}(x)=(2-\sqrt{2}C_{+})\left(\dfrac{1}{3-2\sqrt{2}C_{+}}\right)^{|x|}\overline{\mu}^{(1)}(0)+(2-\sqrt{2}C_{-})\left(\dfrac{1}{3-2\sqrt{2}C_{-}}\right)^{|x|}\overline{\mu}^{(2)}(0),\]
\noindent
where
\[\overline{\mu}^{(1)}(0)=\dfrac{(1-\sqrt{2}C_{+})^{2}}{(3-2\sqrt{2}C_{+})^{2}}|\alpha+i\beta|^{2}I_{(0,3/4)}(\phi),\] 
\noindent
and 
\[\overline{\mu}^{(2)}(0)=\dfrac{(1-\sqrt{2}C_{-})^{2}}{(3-2\sqrt{2}C_{-})^{2}}|\alpha-i\beta|^{2}I_{(1/4,1)}(\phi),\]
with
\[
\left\{ \begin{array}{l}
C_{+}=\cos\left(2\pi\phi+\dfrac{\pi}{4}\right)=\dfrac{1}{\sqrt{2}}\{\cos(2\pi\phi)-\sin(2\pi\phi)\},\\
C_{-}=\cos\left(2\pi\phi-\dfrac{\pi}{4}\right)=\dfrac{1}{\sqrt{2}}\{\cos(2\pi\phi)+\sin(2\pi\phi)\}.
\end{array} \right.\]
\end{enumerate}
\end{theo}
\par\indent
\par\indent
We emphasize that the time-averaged limit measure is symmetric for the origin and 
localization heavily depends on the choice of the initial coin state $\varphi$ and parameter $\phi$. 
For instance, when $\alpha=i\beta$ and $\phi\in(3/4,1)$, we see $\overline{\mu}_{\infty}(x)=0\;(x\in\mathbb{Z})$ holds.
When $\alpha=-i\beta$ and $\phi\in(0,1/4)$, we also have $\overline{\mu}_{\infty}(x)=0\;(x\in\mathbb{Z})$.
Moreover, our results imply that the stationary measure given by \cite{watanabe} stated below is a special case for the time-averaged limit measure.
The proof of Theorem \ref{main-result2} is given in Sect. $6$.

Here we consider the relation between the time-averaged and stationary measures.
First, we present the statioary measure for the Wojcik model in Theorem $2$ of Ref. \cite{watanabe} as follows:
\par\indent
\par\noindent
\begin{theo}
\[\mu(x)=\|\Psi(x)\|^{2}=2|\alpha|^{2}|\theta_{s}|^{2|x|}
\times\left\{\begin{array}{ll}\Gamma(\phi)&(x\neq0),\\
1&(x=0),\end{array}\right.\]
where
\begin{align*}
\Gamma(\phi)=\left\{ \begin{array}{ll}
2-\cos(2\pi\phi)-\sin(2\pi\phi)&(\beta=i\alpha),\\
2-\cos(2\pi\phi)+\sin(2\pi\phi)&(\beta=-i\alpha),
\end{array} \right.
\end{align*}
and
\begin{align}|\theta_{s}|^{2}=\left\{ \begin{array}{ll}
\dfrac{1}{3-2\cos(2\pi\phi)-2\sin(2\pi\phi)} & (\beta=i\alpha), \\
&\\
\dfrac{1}{3-2\cos(2\pi\phi)+2\sin(2\pi\phi)} & (\beta=-i\alpha). 
\end{array} \right.\end{align}
\end{theo}
\par\indent
\par\noindent
We should note that \[\dfrac{1}{3-2\sqrt{2}C_{+}}=\dfrac{1}{3-2\cos(2\pi\phi)-\sin(2\pi\phi)}\] and 
\[\dfrac{1}{3-2\sqrt{2}C_{-}}=\dfrac{1}{3-2\cos(2\pi\phi)+2\sin(2\pi\phi)}\] 
in Theorem \ref{main-result2} agree with $|\theta_{s}|^{2}$:
\begin{align*}|\theta_{s}|^{2}=\left\{ \begin{array}{ll}
\dfrac{1}{3-2\cos(2\pi\phi)-2\sin(2\pi\phi)} & (\beta=i\alpha), \\
&\\
\dfrac{1}{3-2\cos(2\pi\phi)+2\sin(2\pi\phi)} & (\beta=-i\alpha). 
\end{array} \right.\end{align*}
\noindent
From now on, we consider the two cases.
When $\alpha=1/\sqrt{2}$ and $\beta=i/\sqrt{2}$ for instance, we have $\alpha+i\beta=0$ and $\alpha-i\beta=\sqrt{2}$.
Then we have the time-averaged limit measure as follows.
\begin{align}\overline{\mu}_{\infty}(x)=\left\{\begin{array}{ll}
\dfrac{2(1-\sqrt{2}C_{-})^{2}}{(3-2\sqrt{2}C_{-})^{2}}I_{(1/4,1)}(\phi)&(x=0), \\\
(2-\sqrt{2}C_{-})\left(\dfrac{1}{3-2\sqrt{2}C_{-}}\right)^{|x|}\overline{\mu}_{\infty}(0)&(x\neq0). 
\end{array} \right.\label{(6)}
\end{align}
On the other hand, Theorem $2$ in Ref. \cite{watanabe} gives the stationary measure as
\begin{align}
\mu_{\infty}(x)=\left\{\begin{array}{ll}
|c|^{2}&(x=0),\\
(2-\sqrt{2}C_{-})|c|^{2}\left(\dfrac{1}{3-2\sqrt{2}C_{-}}\right)^{|x|}&(x\neq0).\end{array} \right.\label{(7)}
\end{align}
Equations (\ref{(6)}) and (\ref{(7)}) suggest that when $|c|^{2}=2(1-\sqrt{2}C_{-})^{2}/(3-2\sqrt{2}C_{-})^{2}$, 
then the time-averaged limit measure coincides with the stationary measure.

Next, when $\alpha=1/\sqrt{2}$ and $\beta=-i/\sqrt{2}$, we have $\alpha+i\beta=\sqrt{2}$ and $\alpha-i\beta=0$.
Then we obtain the time-averaged limit measure as follows.
\begin{align}\overline{\mu}_{\infty}(x)=\left\{\begin{array}{ll}
\dfrac{2(1-\sqrt{2}C_{+})^{2}}{(3-2\sqrt{2}C_{+})^{2}}I_{(0,3/4)}(\phi)&(x=0), \\\
(2-\sqrt{2}C_{+})\left(\dfrac{1}{3-2\sqrt{2}C_{+}}\right)^{|x|}\overline{\mu}_{\infty}(0)&(x\neq0). 
\end{array} \right.\label{(6')}
\end{align}
On the other hand, Theorem $2$ in Ref. \cite{watanabe} gives the stationary measure.
\begin{align}
\mu_{\infty}(x)=\left\{\begin{array}{ll}
|c|^{2}&(x=0),\\
(2-\sqrt{2}C_{+})|c|^{2}\left(\dfrac{1}{3-2\sqrt{2}C_{+}}\right)^{|x|}&(x\neq0).\end{array} \right.\label{(7')}
\end{align}
Equations (\ref{(6')}) and (\ref{(7')}) suggest that when $|c|^{2}=2(1-\sqrt{2}C_{+})^{2}/(3-2\sqrt{2}C_{+})^{2}$,  
then the time-averaged limit measure also coincides with the stationary measure.
\par\indent
\par\noindent

\section{Result via the CGMV method}
We can derive the time-averaged limit measure at the origin $\overline{\mu}_{\infty}(0)$ also from the CGMV method \cite{cantero}.
From now on, we use the same expressions as in Ref. \cite{cantero}.
Applying the CGMV method to the Wojcik model, we have
\begin{eqnarray*}
a=\dfrac{i}{\sqrt{2}}e^{-2\pi\phi i},\quad b=\dfrac{i}{\sqrt{2}},\quad w=e^{2\pi\phi i},\quad 
\zeta_{\pm}(b)=\pm\dfrac{1}{\sqrt{2}}+
\dfrac{i}{\sqrt{2}}.\label{(25)}
\end{eqnarray*}
As condition ${\mathcal M}_{+}$, we see that the following inequality holds.
\begin{align*}
\dfrac{1}{4}<\phi<1.
\end{align*}
On the other hand, as condition $\mathcal{M}_{-}$, we obtain
\begin{align*}
0<\phi<\dfrac{3}{4}.
\end{align*}
Moreover, we have
\begin{eqnarray*}
\sigma_{1}=0,\;\sigma_{2}=\pi,\quad\sigma=\sigma_{1}+\sigma_{2}=\pi,\quad\theta=\dfrac{\sigma}{2}=\dfrac{\pi}{2},\\
\tau_{1}=2\pi\phi,\quad\tau_{2}=2\pi\phi+\pi,\quad\tau=\tau_{1}+\tau_{2}=4\pi\phi+\pi,\label{(28)}
\end{eqnarray*}
and
\begin{eqnarray*}
C_{+}=\dfrac{1}{\sqrt{2}}(C+S),\quad C_{-}=\dfrac{1}{\sqrt{2}}(C-S),\label{(29)}
\end{eqnarray*}
with
\begin{align*}
C=\cos(2\pi\phi),\quad S=\sin(2\pi\phi).
\end{align*}
Conditions ${\mathcal M}_{+}$ and ${\bf \mathcal{M}}$ imply $1/4<\phi<1$ and $0<\phi<3/4$, respectively.
According to the CGMV method, we get
\begin{align}
\lim_{n\to\infty}P^{(0)}_{\alpha,\beta}(2n)=\dfrac{1}{2}\left(1-\dfrac{\rho_{a}^{2}}{|\zeta_{\pm}(b)-a|^{2}}\right)^{2}
\left\{
1\mp\dfrac{(|\hat{\alpha}|^{2}-|\hat{\beta}|^{2})\Re b+2\rho_{b}\Re(\overline{\omega\hat{\alpha}}\hat{\beta})}
{\sqrt{1-\Im^{2}b}}
\right\}
,\label{300}\end{align}
where $P^{(0)}_{\alpha,\beta}(2n)$ is the probability that the walker return to the origin at time $2n$
 with the initial coin state $\varphi={}^T\![\alpha,\beta]$ where $\alpha,\beta\in\mathbb{C}$ and $|\alpha|^{2}+|\beta|^{2}=1$.
Here,
\begin{eqnarray*}
\rho_{a}=\dfrac{1}{\sqrt{2}},\quad|\zeta_{\pm}(b)-a|^{2}=\dfrac{1}{2}(3-2(C\pm S))=\dfrac{1}{2}(3-2\sqrt{2}C_{\pm}),\nonumber\\
\Im b=\dfrac{1}{\sqrt{2}},\quad\Re b=0,\quad\rho_{b}=\dfrac{1}{\sqrt{2}},\quad\hat{\alpha}=\hat{\lambda}^{(1)}_{0}\alpha=\alpha,\nonumber\\
\hat{\beta}=\hat{\lambda}^{(2)}_{1}\beta=e^{i\left((\sigma_{2}-\sigma_{1})/2+\tau_{2}-\sigma_{2}\right)}\beta=e^{\pi i/2}\omega=i\omega.
\end{eqnarray*}
Therefore, Eq. (\ref{300}) becomes
\begin{eqnarray*}
\lim_{n\to\infty}P^{(0)}_{\alpha,\beta}(2n)=\left\{\begin{array}{ll}
\dfrac{2(1-\sqrt{2}C_{\pm})^{2}}{(3-2\sqrt{2}C_{\pm})^{2}}|\alpha-i\beta|^{2}&(\phi\in(1/4,1)),\\
&\\
\dfrac{2(1-\sqrt{2}C_{\pm})^{2}}{(3-2\sqrt{2}C_{\pm})^{2}}|\alpha+i\beta|^{2}&(\phi\in(0,3/4)).\\
\end{array} \right.\label{(33)}
\end{eqnarray*} 
Thus, we obtain the time-averaged limit measure $\overline{\mu}_{\infty}(0)$ as follows.
\begin{eqnarray*}
\overline{\mu}_{\infty}(0)\!\!\!\!\!&=&\!\!\!\!\!\dfrac{1}{2}\lim_{n\to\infty}P^{(0)}_{\alpha,\beta}(2n)\nonumber\\
\!\!\!\!\!&=&\!\!\!\!\!\left(\dfrac{1-\sqrt{2}C_{+}}{3-2\sqrt{2}C_{+}}\right)^{2}|\alpha-i\beta|^{2}I_{(1/4,1)}(\phi)+
\left(\dfrac{1-\sqrt{2}C_{-}}{3-2\sqrt{2}C_{-}}\right)^{2}|\alpha+i\beta|^{2}I_{(0,3/4)}(\phi),\label{(34)}
\end{eqnarray*}
which agrees with our result.
\par\indent
\section{Proof of Lemma \ref{wojcik-lemma1}}
\label{wojcik-lemma1-proof}
At first, we consider the case of the Hadamard walk on $\mathbb{Z}_{\geq}=\{0,1,2,\cdots\}$ starting at $m\;(\geq1)$.
\begin{align}
U_x = H = \frac{1}{\sqrt{2}}
\left[
\begin{array}{cc}
1 & 1 \\
1 & -1
\end{array}
\right]\quad(x\geq0),
\label{wojcik-lemma1-proof1}
\end{align}
which can be devided into $P_{x}$ and $Q_{x}$ as 
\begin{align}U_{x}=P_{x}+Q_{x}\quad(x\geq1),\label{wojcik-lemma1-proof2}\end{align}
where
\begin{align*}
P_x = P = \frac{1}{\sqrt{2}}
\left[
\begin{array}{cc}
1 & 1 \\
0 & 0
\end{array}
\right],
\quad
Q_x = Q = \frac{1}{\sqrt{2}}
\left[
\begin{array}{cc}
0 & 0 \\
1 & -1
\end{array}
\right].
\end{align*}
Next let $\Xi^{(\infty,m)}_n$ be the sum of all the passages starting at $m\;(\geq1)$ 
and arrive at the origin at time $n$ for the first time. 
For instance, we have
\begin{align*} 
\Xi^{(\infty,1)} _5  = P^2 Q P Q + P^3 Q^2.
\end{align*} 
Here we introduce $R$ and $S$ as 
\begin{align*}
R =
\frac{1}{\sqrt{2}}
\left[
\begin{array}{cc}
1 & -1 \\
0 & 0 
\end{array}
\right], 
\quad
S=
\frac{1}{\sqrt{2}}
\left[
\begin{array}{cc}
0 & 0 \\
1 & 1 
\end{array}
\right].
\end{align*}
\par\indent
\par\noindent
We should remark that the matrices $P,\;Q,\;R,$ and $S$ become the orthonormal basis of the vector space consisting of $2\times2$ matrices 
for the inner product $\langle A|B\rangle=tr(A^{\ast}B)$. Therefore, $\Xi^{(\infty,m)} _n$ ($m \geq 1$) can be written in terms of $P,Q,R,$ and $S$ uniquely as
\begin{align} 
\Xi^{(\infty,m)} _n = p^{(\infty,m)} _n P + q^{(\infty,m)} _n Q + r^{(\infty,m)} _n R + s^{(\infty,m)} _n S,
\label{wojcik-lemma1-proof3}
\end{align}
where $p^{(\infty,m)} _n,\; q^{(\infty,m)} _n,\;r^{(\infty,m)} _n,\;s^{(\infty,m)} _n \in\mathbb{C}$.
Now, by the definition of $\Xi^{(\infty,m)} _n$, we have 
\begin{align*} 
\Xi^{(\infty,m)} _n = \Xi^{(\infty,m-1)} _{n-1} P + \Xi^{(\infty,m+1)} _{n-1} Q,\end{align*}
which yields the coefficients as follows.
\begin{align*}
p^{(\infty,m)} _n 
&= \frac{1}{\sqrt{2}} \> p^{(\infty,m-1)} _{n-1} + \frac{1}{\sqrt{2}} \> r^{(\infty,m-1)} _{n-1}, 
\\  
q^{(\infty,m)} _n 
&= - \frac{1}{\sqrt{2}} \> q^{(\infty,m+1)} _{n-1} + \frac{1}{\sqrt{2}} \> s^{(\infty,m+1)} _{n-1}, 
\\ 
r^{(\infty,m)} _n 
&= \frac{1}{\sqrt{2}} \> p^{(\infty,m+1)} _{n-1} - \frac{1}{\sqrt{2}} \> r^{(\infty,m+1)} _{n-1}, 
\\ 
s^{(\infty,m)} _n 
&= \frac{1}{\sqrt{2}} \> q^{(\infty,m-1)} _{n-1} + \frac{1}{\sqrt{2}} \> s^{(\infty,m-1)} _{n-1}. 
\end{align*}
Here, according to the definition of $\Xi^{(\infty,m)} _n$, we see that the walker arrives at the origin with 
the final step to the left, and we have only two types of the passages satisfying the condition, that is, $P \ldots P$ or $P \ldots Q$,
which leads to $q^{(\infty,m)} _n=s^{(\infty,m)} _n=0 \;(n \ge 1)$.
To compute $p^{(\infty,m)} _n$ and $r^{(\infty,m)} _n$, we introduce the generating functions  in the following.
\begin{align}
p^{(\infty,m)}  (z) = \sum_{n=1} ^{\infty} p^{(\infty,m)} _n z^n, \quad r^{(\infty,m)}  (z) = \sum_{n=1} ^{\infty}.
r^{(\infty,m)} _n z^n.\label{wojcik-lemma1-proof4}
\end{align}
Equation (\ref{wojcik-lemma1-proof4}) yields
\begin{align}
p^{(\infty,m)} (z) &= \frac{z}{\sqrt{2}} \> p^{(\infty,m-1)} (z) + \frac{z}{\sqrt{2}} \> r^{(\infty,m-1)} (z), \label{wojcik-lemma1-proof5}\\
r^{(\infty,m)} (z) &= \frac{z}{\sqrt{2}} \> p^{(\infty,m+1)} (z) - \frac{z}{\sqrt{2}} \> r^{(\infty,m+1)} (z).\label{wojcik-lemma1-proof6} \end{align} 
Equations (\ref{wojcik-lemma1-proof5}) and (\ref{wojcik-lemma1-proof6}) give the recurrence formula for $p^{(\infty,m)} (z)$ and $r^{(\infty,m)}  (z)$ 
as   
\begin{align} 
p^{(\infty,m+2)} (z) + \sqrt{2} \> \left( {1 \over z} - z \right) p^{(\infty,m+1)} (z) - p^{(\infty,m)} (z) &= 0,
\label{wojcik-lemma1-proof7}\\
r^{(\infty,m+2)} (z) + \sqrt{2} \> \left( {1 \over z} - z \right) r^{(\infty,m+1)} (z) - r^{(\infty,m)} (z) &= 0.
\label{wojcik-lemma1-proof8}
\end{align} 
Thus, we see that $p^{(\infty,m)} (z)$ and $r^{(\infty,m)}$ satisfy the same recurrence formula, and
the characteristic polynomial has the two solutions:
\begin{align*}\lambda_{\pm} =\frac{-1 + z^2 \pm \sqrt{1+z^4}}{\sqrt{2}z}.\end{align*}

Next the definition of $\Xi^{(\infty,1)} _n$ gives $p^{(\infty,1)}_n = 0 \> (n \ge 2)$ and $p_1 ^{(\infty,1)} =1$, 
and we have
 $p^{(\infty,1)}(z) = z$. Moreover, noting $\lim_{m \to \infty} p^{(\infty,m)} (z) < \infty$, we obtain
\begin{align}
p^{(\infty,m)} (z) = z \lambda_ +^{m-1}, \quad r^{(\infty,m)} (z) = \frac{-1+\sqrt{1+z^4}}{z} \lambda_+^{m-1}.
\label{wojcik-lemma1-proof9}\end{align}
Hence we get
\begin{align}
r^{(\infty,1)} (z) = \frac{-1+\sqrt{1+ z^4}}{z},
\label{wojcik-lemma1-proof10}\end{align} 
for $m=1$.
In a similar way, we obtain
\begin{align*}
q^{(-\infty,m)} (z) = z \lambda_- ^{m+1}, \quad s^{(-\infty,m)} (z) = \frac{1-\sqrt{1+ z^4}}{z} \lambda_-^{m+1},
\end{align*}
for the Hadamard walk on $\mathbb{Z}_{\leq} = \{ 0, -1, -2, \ldots \}$ starting at $m (\le -1)$.
Therefore, we have
\begin{align}
s^{(- \infty, -1)} (z) = \frac{1 - \sqrt{1 + z^4}}{z}\label{wojcik-lemma1-proof11}
\end{align} 
for $m=-1$.
Here we shoud note that 
\begin{align}r_{n}^{(\infty,1)} + s_{n}^{(-\infty,-1)} = 0\label{wojcik-lemma1-proof12}\end{align}
holds for $n\geq1$.
Next we put 
\begin{align}\Xi_n^{+} = \Xi_{n-1}^{(\infty, 1)} Q_0,\quad\Xi_n^{-} = \Xi_{n-1}^{(-\infty, -1)} P_0,\label{wojcik-lemma1-proof13}\end{align}
where
\begin{align*}
P_0 = \omega P = \frac{\omega}{\sqrt{2}}
\left[
\begin{array}{cc}
1 & 1 \\
0 & 0 
\end{array}
\right], 
\quad
Q_0 = \omega Q = \frac{\omega}{\sqrt{2}}
\left[
\begin{array}{cc}
0 & 0 \\
1 & -1
\end{array}
\right].
\end{align*}
We notice that $\Xi_n^{+}$ ($\Xi_n^{-}$) is the sum of all the passages that the walker arrives at the origin 
 at time $n$ for the first time in $\mathbb{Z}_{\geq}$ ($\mathbb{Z}_{\le}$).
Thus, we have
\par\indent
\par\noindent
\begin{lemma}
\label{wojcik-lemma4}
$(i)$ If $n\in\mathbb{N}$ is even and $n\geq4$, we have
\begin{align*}
\Xi_n^{+} 
&= r^{(\infty,1)}_{n-1} \> R Q_0 
= \frac{\omega r^{(\infty,1)}_{n-1}}{2} 
\left[
\begin{array}{cc}
-1 & 1 \\
0 & 0
\end{array}
\right],
\\
\Xi_n^{-} 
&= s^{(-\infty,-1)}_{n-1} \> S P_0 
= \frac{\omega s^{(-\infty,-1)}_{n-1}}{2} 
\left[
\begin{array}{cc}
0 & 0 \\
1 & 1
\end{array}
\right],
\end{align*}
where 
\begin{align*}
\sum_{n=1}^{\infty} \> r_{n}^{(\infty,1)} z^n = \frac{-1 + \sqrt{1 + z^4}}{z}, \quad \sum_{n=1}^{\infty} \> 
s_{n}^{(-\infty,-1)} z^n = \frac{1 - \sqrt{1 + z^4}}{z}.
\end{align*}
$(ii)$ 
\begin{align*}
\Xi_2^{+} 
= P Q_0 
= \frac{-\omega}{2} 
\left[
\begin{array}{cc}
-1 & 1 \\
0 & 0
\end{array}
\right], \qquad
\Xi_2^{-} 
= Q P_0 
= \frac{\omega}{2} 
\left[
\begin{array}{cc}
0 & 0 \\
1 & 1
\end{array}
\right].
\end{align*}
$(iii)$ If $n\in\mathbb{N}$ is odd, we have
\begin{align*}
\Xi_n^{+} = \Xi_n^{-} 
= 
\left[
\begin{array}{cc}
0 & 0 \\
0 & 0
\end{array}
\right].
\end{align*}
\end{lemma}

Here if we put $\Xi_n^{\ast} = \Xi_n^{+} + \Xi_n^{-}$, then
Lemma \ref{wojcik-lemma2} and 
\begin{align}s_{n}^{(-\infty,-1)} = - r_{n}^{(\infty,1)} \> \quad(n \ge 1)\label{wojcik-lemma1-proof14}\end{align}
give
\begin{align*}
\Xi^{\ast}_n
=\frac{\omega r^{\ast}_{n-1}}{2} \> 
\left[
\begin{array}{cc}
-1 & 1 \\
-1 & -1
\end{array}
\right],
\end{align*}
where
\begin{align*}
r_n^{\ast} = 
\left\{
\begin{array}{cl}
\displaystyle{(-1)^{m-1} \> \frac{(2m-1)!}{2^{2m-1} (m-1)! m!}} & \quad \mbox{($n=4m-1 \>\> m \ge 1$),} \\
0 & \quad \mbox{($n \not= 4m-1, \> n \ge 2, \>\> m \ge 1$),} \\
-1 & \quad \mbox{($n =1$).}
\end{array}
\right.
\end{align*}
Then, we see
\begin{align*}
r_1^{\ast} 
&= -1, \> r_2^{\ast} = 0, \> r_3^{\ast} = 1/2,  \> r_4^{\ast} = r_5^{\ast} = r_6^{\ast} = 0, \> \\
r_7^{\ast} 
&= -1/8, \> r_8^{\ast} = r_9^{\ast} = r_{10}^{\ast} = 0, \ldots.   
\end{align*}
The generating function of $r_{n}^{\ast}$ is given by
\begin{align*}
\sum_{n=1}^{\infty} \> r_{n}^{\ast} z^n = \frac{-1 - z^2 + \sqrt{1 + z^4}}{z}.
\end{align*}
From the definition of $\Xi^{\ast}_{n}$, we see
\begin{align*}
\Psi_{2n} (0) 
= \sum_{k=1}^n \sum_{\scriptstyle (a_1, \ldots , a_k) \in (\mathbb{Z}_{>})^k : \atop \scriptstyle a_1 + \cdots + a_k =n} \left( \prod_{j=1}^k \Xi^{\ast}_{2 a_j} \right) \> \varphi. 
\end{align*}
where $\mathbb{Z}_{>} = \{1,2, \ldots \}$. Moreover, the following relation holds for $\eta =1,-1$:
\begin{align*}
\left[
\begin{array}{cc}
-1 & 1 \\
-1 & -1
\end{array}
\right]^k \> \frac{1}{\sqrt{2}}
\left[
\begin{array}{cc}
1 \\
\eta i
\end{array}
\right]
= 
\frac{(-1+\eta i)^k}{\sqrt{2}} 
\left[
\begin{array}{cc}
1 \\
\eta i
\end{array}
\right].
\end{align*}
From the case (iii) in Lemma \ref{wojcik-lemma4}, we have
\begin{align}
\left( \prod_{j=1}^k \Xi^{\ast}_{2 a_j} \right) \> \varphi
= 
\left( \prod_{j=1}^k r^{\ast}_{2 a_j-1} \right) \> \left( \frac{\omega}{2} \right)^k \> 
\left[
\begin{array}{cc}
-1 & 1 \\
-1 & -1
\end{array}
\right]^k \> 
\frac{1}{\sqrt{2}}
\left[
\begin{array}{cc}
1 \\
\eta i
\end{array}
\right],\label{wojcik-lemma1-proof15}
\end{align}
and we arrive at the desired conclusion.

\section{Proof of Theorem \ref{main-result2}} 
Taking advantage of the generating function for the weight of the passages, we give the proof of Theorem 
\ref{main-result2} 
by the following $3$ steps. \\
\par\noindent
{\bf $(1)$ Facts}\\
\par\indent
We begin with introducing some notations. Let the QW be a space-inhomogeneous model on the line defined by $2\times2$ unitary matrices
\begin{align}
U_{x}=\begin{bmatrix}
a_{x}&b_{x}\\
c_{x}&d_{x}
\end{bmatrix}\quad(x\in \mathbb{Z}).\label{(8)}
\end{align}
The subscript $x$ expresses the position.
We should recall that $U_{x}$ can be devided into two parts as 
\[U_{x}=P_{x}+Q_{x},\]
where
\[ 
P_{x}=\begin{bmatrix} 
a_{x} & b_{x} \\
0 & 0 
\end{bmatrix},\quad
Q_{x}=\begin{bmatrix}
0 & 0 \\
c_{x} & d_{x}
\end{bmatrix}.\]
Konno et al. \cite{segawa} showed the following key result in the proof of Theorem $2$.
\par\indent
\begin{prop}
\label{prf2.1}
The time-averaged limit measure defined by Eq. (\ref{(2)}) in Sect. $2$ is expressed by
\[\overline{\mu}_{\infty}(x)=\sum_{\theta_{s}}\|{\rm Res}(\tilde{\Xi}_{x}(z):z=e^{i\theta_{s}})\varphi\|^{2},\]
where $\tilde{\Xi}_{x}(z)=\sum_{n\geq0}\Xi_{n}(x)z^{n}$ and $\{e^{i\theta_{s}}\}$ is the set of the singular points of $\tilde{\Xi}_{x}(z)$.
\end{prop}
\par\indent
\par\noindent
Konno et al. \cite{segawa} also presened the key expressions of $\tilde{\Xi}_{x}(z)$ as follows.

\par\indent
\begin{lemma}
\label{prf2.2}
 Let $\Delta_{x}$ express the determinant of $U_{x}$. 
We assume that $a_{x},d_{x}\neq 0$ for all $x\in\mathbb{Z}$.
\begin{enumerate}
\item If $x=0$, we have 
\[\tilde{\Xi}_{0}(z)=\dfrac{1}{1-\sqrt{2}\omega\tilde{f}_{0}(z)+\omega^{2}\{\tilde{f}_{0}(z)\}^{2}}
\begin{bmatrix} 
1-\dfrac{e^{2\pi i\phi}}{\sqrt{2}}\tilde{f}_{0}(z) & -\dfrac{e^{2\phi i\phi}}{\sqrt{2}}\tilde{f}_{0}(z)\\
&\\
\dfrac{e^{2\pi i\phi}}{\sqrt{2}}\tilde{f}_{0}(z) & 1-\dfrac{e^{2\pi i\phi}}{\sqrt{2}}\tilde{f}_{0}(z)\\
\end{bmatrix}.\]

\item If $|x|\geq 1$, we have
\[
\tilde{\Xi}_{x}(z)=\left\{\begin{array}{ll}
(\tilde{\lambda}_{x}(z))^{x-1}
\left[
    \begin{array}{c}
      \tilde{\lambda}_{x}(z)\tilde{f}_{x}(z)\\
      z \\
    \end{array}
  \right]\left[\dfrac{e^{2\pi i\phi}}{\sqrt{2}},-\dfrac{e^{2\pi i\phi}}{\sqrt{2}}\right]\tilde{\Xi}_{0}(z) & (x\geq 1), \\
  &\\
(-\tilde{\lambda}_{x}(z))^{|x|-1}
\left[
    \begin{array}{c}
      z \\
      -\tilde{\lambda}_{x}(z)\tilde{f}_{x}(z) \\
    \end{array}
  \right]\left[\dfrac{e^{2\pi i\phi}}{\sqrt{2}},\dfrac{e^{2\pi i\phi}}{\sqrt{2}}\right]\tilde{\Xi}_{0}(z) & (x\leq -1), \\
\end{array} \right.\]
\end{enumerate}
where $\tilde{\lambda}_{x}(z)=\dfrac{z}{\tilde{f}_{x}(z)-\sqrt{2}}$. 
Here $\tilde{f}_{x}(z)$ satisfies the following quadratic equation.
\begin{align}(\tilde{f}_{x}(z))^{2}-\sqrt{2}(1+z^{2})\tilde{f}_{x}(z)+z^{2}=0.\label{(Takako1)}\end{align}
\end{lemma}
Now noting
\[\tilde{f}^{(\pm)}_{x}(z)=\tilde{f}^{(\pm)}_{0}(z),\;\tilde{f}^{(+)}_{x}(z)=\tilde{f}^{(-)}_{x}(z),
\;\tilde{\lambda}^{(\pm)}_{x}(z)=\tilde{\lambda}^{(\pm)}_{0}(z),\]
we put $\tilde{f}(z)$ and $\tilde{\lambda}^{(\pm)}(z)$, respectivelly.
Next, we give the expressions for $\tilde{f}(z)$ in terms of $\theta$ which link to the singular points for 
$\tilde{\Xi}_{x}(z)$.
\par\indent
\begin{lemma}
\label{prf2.3}
For $z=e^{i\theta}\;(\theta\in\mathbb{R})$, we have 
\begin{align}\tilde{f}(e^{i\theta})=e^{i(\theta+\tilde{\phi}(\theta))},\label{(9)}\end{align}
where $\tilde{\phi}(\theta)$ is defined by 
\begin{align}\left\{
\begin{array}{l}
\sin\tilde{\phi}(\theta)=\operatorname{sgn}(\sin\theta)\sqrt{2\sin\theta^{2}-1},\\
\cos\tilde{\phi}(\theta)=\sqrt{2}\cos\theta.
\end{array}
\right.\label{(10)}\end{align} 
\end{lemma}
{\bf Proof.}
Noting $\tilde{f}(0)=0$, Eq. (\ref{(Takako1)}) gives
\begin{align}
\tilde{f}(z)=\dfrac{z^{2}+1-\sqrt{z^{4}+1}}{\sqrt{2}}.\label{(Takako2)}
\end{align}
Putting $z=e^{i\theta}$, taking advantage of the explicit expressions for $\tilde{f}(z)$ and $\tilde{\lambda}(z)$ in Ref. \cite{segawa},
we obtain
\begin{align*}
\tilde{f}(e^{i\theta})=e^{i\theta}(\sqrt{2}\cos\theta+i\operatorname{sgn}(\sin\theta)\sqrt{1-2\cos^{2}\theta}),
\end{align*}
which completes the proof.
\par\indent

Now the singular points of $\tilde{\Xi}_{x}(z)$ are given as follows.
\par\indent
\begin{lemma}
\label{prf2.4}
Let
\[e^{i\theta^{(\pm)}_{1}}=\pm\left(\dfrac{\sin C_{+}}{\sqrt{3-2\sqrt{2}\sin C_{+}}}+\dfrac{\cos C_{+}-\sqrt{2}}{\sqrt{3-2\sqrt{2}\cos C_{+}}}i\right),\quad
e^{i\theta^{(\pm)}_{2}}=\pm\left(\dfrac{\sin C_{-}}{\sqrt{3-2\sqrt{2}\sin C_{-}}}+\dfrac{\cos C_{-}-\sqrt{2}}{\sqrt{3-2\sqrt{2}\cos C_{-}}}i\right),\]
where
\[
\left\{ \begin{array}{l}
C_{+}=\cos\left(2\pi\phi+\dfrac{\pi}{4}\right)=\dfrac{1}{\sqrt{2}}\{\cos(2\pi\phi)-\sin(2\pi\phi)\},\\
C_{-}=\cos\left(2\pi\phi-\dfrac{\pi}{4}\right)=\dfrac{1}{\sqrt{2}}\{\cos(2\pi\phi)+\sin(2\pi\phi)\}
.\end{array} \right.\]
Here, the set of the singular points for $\;\tilde{\Xi}_{x}(z)$ with $|z|=1$, are given by
\[B=\{e^{i\theta_{1}^{(\pm)}}, e^{i\theta_{2}^{(\pm)}}\}.\]
\end{lemma}
We give the proof of Lemma \ref{prf2.4} in Appendix B.\\ 

Next we derive the residues of $\;\tilde{\Xi}_{x}(z)$ at each singular point. From now on, we put $\tilde{\Lambda}_{0}(z)\equiv1-\sqrt{2}\omega\tilde{f}(z)+\omega^{2}\tilde{f}^{2}(z)^{2}$,
where $\omega=e^{2\pi i\phi}$ with $\phi\in(0,1)$. Remark that each singular point $z\in B$ is derived from the solution for 
\[\tilde{\Lambda}_{0}(z)=0.\]
\par\noindent
{\bf $(2)$ Proof for $x=0$ case.}\\
\par\noindent
First of all, we present the proof of Theorem \ref{main-result2} for $x=0$ case.
Here Proposition $1$ gives 
\begin{align}
\overline{\mu}_{\infty}(0)=\sum_{\theta_{s}}\|{\rm Res}(\tilde{\Xi}_{0}(z)\varphi:z=e^{i\theta_{s}})\|^{2},\label{(11)}
\end{align}
where $\{e^{i\theta_{s}}\}$ is the set of the singular points for $\tilde{\Xi}_{0}(z)$.
According to Lemma $1$, we have
\[\tilde{\Xi}_{0}(z)\varphi=\dfrac{1}{\tilde{\Lambda}_{0}(z)}
\begin{bmatrix}\alpha\left(1-\dfrac{e^{2\pi i\phi}}{\sqrt{2}}\tilde{f}(z)\right)
-\dfrac{\beta e^{2\pi i\phi}}{\sqrt{2}}\tilde{f}(z)\\\ 
 & \\
\dfrac{\alpha e^{2\pi i\phi}}{\sqrt{2}}\tilde{f}(z)
+\beta\left(1-\dfrac{e^{2\pi i\phi}}{\sqrt{2}}\tilde{f}(z)\right)\end{bmatrix},\]
where we put $\alpha=\Psi^{L}(0)$ and $\beta=\Psi^{R}(0)$ in short.
Then the square norm of the residue is defined by
\begin{eqnarray}\|{\rm Res}(\tilde{\Xi}_{0}(z)\varphi:z=e^{i\theta_{s}})\|^{2}
=\left|{\rm Res}\left(\dfrac{\alpha\left(1-\dfrac{e^{2\pi i\phi}}{\sqrt{2}}\tilde{f}(z)\right)
-\dfrac{\beta e^{2\pi i\phi}}{\sqrt{2}}\tilde{f}(z)}{\tilde{\Lambda}_{0}}:z=e^{i\theta_{s}}\right)\right|^{2}
\nonumber\\
+\left|{\rm Res}\left(\dfrac{\dfrac{\alpha e^{2\pi i\phi}}{\sqrt{2}}\tilde{f}(z)
+\beta\left(1-\dfrac{e^{2\pi i\phi}}{\sqrt{2}}\tilde{f}(z)\right)}{\tilde{\Lambda}_{0}} 
:z=e^{i\theta_{s}}\right)\right|^{2}.\label{(12)}\end{eqnarray}
By the definition of the residue, we have
\[{\rm Res}\left(\dfrac{1}{\tilde{\Lambda}_{0}(z)}:z=e^{i\theta}\right)
=\lim_{z\to e^{i\theta}}\dfrac{z-e^{i\theta}}{\tilde{\Lambda}_{0}(z)},\]
for any $\theta\in\mathbb{R}$.
By expanding $\tilde{\Lambda}_{0}(z)$ around $e^{i\theta_{s}}$, we get
\begin{align}\left|{\rm Res}\left(\frac{1}{\tilde{\Lambda}_{0}(z)}:z=e^{i\theta_{s}}\right)\right|^{2}
=\dfrac{1}{|\tilde{\Lambda}^{'}_{0}(e^{i\theta_{s}})|^{2}}
=\frac{1}{2\left|1+\dfrac{\partial{\tilde{\phi}(\theta)}}{\partial{\theta}}\right|^{2}_{\theta=\theta_{s}}},\label{(13)}\end{align}
where
\[\tilde{\Lambda}^{'}_{0}(z)=\frac{\partial\tilde{\Lambda}_{0}(z)}{\partial z}.\]
\par\indent
Noting Eq.(\ref{(13)}), we compute 
$\|{\rm Res}(\tilde{\Xi}_{0}(z)\varphi:z=e^{i\theta_{s}})\|^{2}$ for each singular point in the following way.\\
Combining Eq.(\ref{(12)}) with Eq.(\ref{(13)}), we see

\begin{align}
\left\{
\begin{array}{l}
\left|{\rm Res}\left(\dfrac{\alpha\left(1-\dfrac{e^{2\pi i\phi}}{\sqrt{2}}\tilde{f}(z)\right)
-\dfrac{\beta e^{2\pi i\phi}}{\sqrt{2}}\tilde{f}(z)}{\tilde{\Lambda}_{0}}:z=e^{i\theta_{s}}\right)\right|^{2}=
\dfrac{\left|\alpha\left(1-\dfrac{e^{2\pi i\phi}}{\sqrt{2}}\tilde{f}(e^{i\theta_{s}})\right)-\dfrac{\beta e^{2\pi i\phi}}{\sqrt{2}}\tilde{f}(e^{i\theta_{s}})\right|^{2}}
{2\left|1+\dfrac{\partial{\tilde{\phi}(\theta)}}{\partial{\theta}}\right|_{\theta=\theta_{s}}^{2}},\\
\\
\left|{\rm Res}\left(\dfrac{\dfrac{\alpha e^{2\pi i\phi}}{\sqrt{2}}\tilde{f}(z)
+\beta\left(1-\dfrac{e^{2\pi i\phi}}{\sqrt{2}}\tilde{f}(z)\right)}{\tilde{\Lambda}^{'}_{0}} :z=e^{i\theta_{s}}\right)\right|^{2}=
\dfrac{\left|\dfrac{\alpha e^{2\pi i\phi}}{\sqrt{2}}\tilde{f}(e^{i\theta_{s}})+\beta\left(1-\dfrac{e^{2\pi i\phi}}
{\sqrt{2}}\tilde{f}(e^{i\theta_{s}})\right)\right|^{2}}{2\left|1+\dfrac{\partial{\tilde{\phi}(\theta)}}{\partial{\theta}}\right|_{\theta=\theta_{s}}^{2}}.\\
\end{array}\right.\label{(14)}\end{align}
\noindent
Substituting the singular points for $\tilde{\Xi}_{0}(z)$ into the right hand side of Eq.(\ref{(14)}), 
$\|{\rm Res}(\tilde{\Xi}_{0}(z)\varphi:z=e^{i\theta_{s}})\|^{2}$ can be written as follows.
\par\indent
\begin{lemma}
\label{prf2.5}
\begin{enumerate}
\item For $e^{i\theta^{(\pm)}_{1}}$, we have
\[\|{\rm Res}(\tilde{\Xi}_{0}(z)\varphi:z=e^{i\theta^{(\pm)}_{1}})\|^{2}=\frac{1}{2}\left|\frac{1-\sqrt{2}C_{+}}{3-2\sqrt{2}C_{+}}\right|^{2}|\alpha+i\beta|^{2}.\]
\item For $e^{i\theta^{(\pm)}_{2}}$, we have
\[\|{\rm Res}(\tilde{\Xi}_{0}(z)\varphi:z=e^{i\theta^{(\pm)}_{2}})\|^{2}=\frac{1}{2}\left|\frac{1-\sqrt{2}C_{-}}
{3-2\sqrt{2}C_{-}}\right|^{2}|\alpha-i\beta|^{2},\]
where
\[
\left\{ \begin{array}{l}
C_{+}=\cos\left(2\pi\phi+\dfrac{\pi}{4}\right)=\dfrac{1}{\sqrt{2}}\{\cos(2\pi\phi)-\sin(2\pi\phi)\},\\
C_{-}=\cos\left(2\pi\phi-\dfrac{\pi}{4}\right)=\dfrac{1}{\sqrt{2}}\{\cos(2\pi\phi)+\sin(2\pi\phi)\}.\end{array} \right.\]
\end{enumerate}
\end{lemma}
\par\noindent
Combining Eq.(\ref{(11)}) with Lemma \ref{prf2.5}, we obtain the desired conclusion for $x=0$ case.\\

\par\noindent
{\bf $(3)$ Proof for $|x|\geq1$ case.}\\
\par\indent
Next we give the proof for $|x|\geq1$ case in Theorem \ref{main-result2} by the same line as that of $x=0$ case.
From Lemma \ref{prf2.2}, we have
\[\tilde{\Xi}_{x}(z)\varphi=\left\{\begin{array}{ll}
\dfrac{e^{2\pi i \phi}(\tilde{\lambda}(z))^{x-1}}{\sqrt{2}\tilde{\Lambda}_{0}(z)}\begin{bmatrix}
\tilde{\lambda}(z)\tilde{f}(z)(\gamma^{L}(z)-\gamma^{R}(z))\\
&\\
z(\gamma^{L}(z)-\gamma^{R}(z))
\end{bmatrix}&(x\geq1),\\
&\\
\dfrac{e^{2\pi i \phi}(-\tilde{\lambda}(z))^{x-1}}{\sqrt{2}\tilde{\Lambda}_{0}(z)}\begin{bmatrix}
z(\gamma^{L}(z)+\gamma^{R}(z))\\
&\\
\tilde{f}(z)(-\tilde{\lambda}(z))(\gamma^{L}(z)+\gamma^{R}(z))
\end{bmatrix}&(x\leq-1),
\end{array} \right.\]
where we put
\[
\left\{ \begin{array}{l}
\gamma^{L}(z)=\alpha\left(1-\dfrac{e^{2\pi i\phi}}{\sqrt{2}}\tilde{f}(z)-\beta\dfrac{e^{2\pi i\phi}}{\sqrt{2}}\tilde{f}(z)\right),\\

\\
\gamma^{R}(z)=\alpha\dfrac{e^{2\pi i\phi}}{\sqrt{2}}\tilde{f}(z)+\beta\left(1-\dfrac{e^{2\pi i\phi}}{\sqrt{2}}\tilde{f}(z)\right).
\end{array} \right.
\]
We should note that the square norm of the residues is defined by
\begin{eqnarray}\|{\rm Res}(\tilde{\Xi}_{x}(z)\varphi:z=e^{i\theta_{s}})\|^{2}
&=\!\left|{\rm Res}\left(\dfrac{e^{2\pi i \phi}(\tilde{\lambda}_{0}(z))^{x}\tilde{f}(z)
(\gamma^{L}(z)-\gamma^{R}(z))}{\sqrt{2}\tilde{\Lambda}_{0}(z)}:z=e^{i\theta_{s}}\right)\right|^{2}
\nonumber\\
&+\!\left|{\rm Res}\left(\dfrac{e^{2\pi i \phi}(\tilde{\lambda}(z))^{x-1}z(\gamma^{L}(z)-\gamma^{R}(z))
}{\sqrt{2}\tilde{\Lambda}_{0}(z)}:z=e^{i\theta_{s}}\right)\right|^{2},\label{(17)}\end{eqnarray}
where $\{e^{i\theta_{s}}\}$ is the set of the singular points for $\tilde{\Xi}_{x}(z).$ According to the definition of the residues, we have
\begin{align}{\rm Res}\left(\dfrac{e^{2\pi i \phi}(\tilde{\lambda}(z))^{x}\tilde{f}(z)(\gamma^{L}(z)-\gamma^{R}(z))}{\sqrt{2}\tilde{\Lambda}_{0}(z)}:z=e^{i\theta_{s}}\right)
=\dfrac{e^{2\pi i \phi}(\tilde{\lambda}(e^{i\theta_{s}}))^{x}\tilde{f}(e^{i\theta})(\gamma^{L}(e^{i\theta_{s}})
-\gamma^{R}(e^{i\theta_{s}}))}{\sqrt{2}\tilde{\Lambda}^{'}_{0}(e^{i\theta_{s}})},\label{(18)}\end{align}
and combining Eq.(\ref{(18)}) with Eq.(\ref{(13)}), we see
\begin{align}\left|{\rm Res}\left(\dfrac{e^{2\pi i \phi}(\tilde{\lambda}(z))^{x}\tilde{f}(z)
(\gamma^{L}(z)-\gamma^{R}(z))}{\sqrt{2}\tilde{\Lambda}_{0}(z)}:z=e^{i\theta_{s}}\right)\right|^{2}
=\dfrac{|\tilde{\lambda}(e^{i\theta_{s}})|^{2x}}{2\left|1+\frac{\partial\tilde{\phi}(\theta)}
{\partial\theta}
\right|_{\theta=\theta_{s}}^{2}}\left|\alpha-\sqrt{2}\alpha e^{2\pi i\phi}\tilde{f}(e^{i\theta_{s}})-\beta\right|^{2}.
\label{(19)}\end{align}
In the same way, we get
\begin{align}\left|{\rm Res}\left(\dfrac{e^{2\pi i \phi}(\tilde{\lambda}(z))^{x-1}z(\gamma^{L}(z)-\gamma^{R}(z))}
{\sqrt{2}\tilde{\Lambda}_{0}(z)}:z=e^{i\theta_{s}}\right)\right|^{2}
=\dfrac{|\tilde{\lambda}(e^{i\theta_{s}})|^{2(x-1)
}}{2\left|1+\frac{\partial\tilde{\phi}(\theta)}{\partial\theta}\right|_{\theta=\theta_{s}}^{2}}
\left|\alpha-\sqrt{2}\alpha e^{2\pi i\phi}\tilde{f}(e^{i\theta_{s}})-\beta\right|^{2}.\label{(20)}\end{align}
Combining Eq.(\ref{(17)}) with Eqs.(\ref{(19)}) and (\ref{(20)}), we obtain
\[\|{\rm Res}(\tilde{\Xi}_{x}(z)\varphi:z=e^{i\theta_{s}})\|^{2}=\frac{|\tilde{\lambda}^{(+)}(e^{i\theta_{s}})|^{2(x-1)}}
{4\left|1+\frac{\partial\tilde{\phi}(\theta)}{\partial\theta}\right|_{\theta=\theta_{s}}^{2}}
(1+|\tilde{\lambda}(e^{i\theta_{s}})|^{2})\left|\alpha-\sqrt{2}\alpha e^{2\pi i\phi}\tilde{f}(e^{i\theta_{s}})-\beta\right|^{2}.\]
Therefore the time-averaged limit measure $\overline{\mu}_{\infty}(x)$ can be written as
\begin{align}\overline{\mu}_{\infty}(x)
=\sum_{\theta_{s}}\frac{|\tilde{\lambda}_{0}(e^{i\theta_{s}})|^{2(x-1)}}{2\left|1+\frac{\partial\tilde{\phi}(\theta)}{\partial\theta}\right|^{2}}
(1+|\tilde{\lambda}(e^{i\theta_{s}})|^{2})\left|\alpha-\sqrt{2}\alpha e^{2\pi i\phi}\tilde{f}(e^{i\theta})-
\beta\right|^{2}.\label{(21)}\end{align}
Next we compute the elements of $\overline{\mu}_{\infty}(x)$ in Eq.(\ref{(21)}) for each singular point of 
$\tilde{\Xi}_{x}$.
\begin{enumerate}
\item $|\tilde{\lambda}(e^{i\theta_{s}})|^{2}\quad(e^{i\theta_{s}}\in B)$:\\
Lemma $1$ gives 
\begin{align}|\tilde{\lambda}(e^{i\theta})|^{2}=3-4\cos^{2}\theta-2\sqrt{2}|\sin\theta|\sqrt{1-2\cos^{2}\theta}.
\label{(22)}\end{align}
Substituting the singular points into Eq.(\ref{(22)}), we obtain $|\tilde{\lambda}(e^{i\theta_{j}^{(\pm)}})|^{2}\;(j=1,2)$ as below.
\begin{enumerate}
\item $e^{i\theta_{1}^{(\pm)}}\in B$ case.\\
\begin{eqnarray}
|\tilde{\lambda}(e^{i\theta^{(\pm)}_{1}})|^{2}=
\dfrac{1}{3-2\sqrt{2}C_{+}}&(C_{+}<1/\sqrt{2}).\label{(23)}\end{eqnarray} 

\item $e^{i\theta_{2}^{(\pm)}}\in B$ case.\\
\begin{eqnarray}
|\tilde{\lambda}(e^{i\theta^{(\pm)}_{2}})|^{2}=
\dfrac{1}{3-2\sqrt{2}C_{-}}&(C_{-}<1/\sqrt{2}).
\label{(25)}\end{eqnarray} 
\end{enumerate}
\item $|\alpha-\sqrt{2}e^{2\pi i\phi}\tilde{f}(e^{i\theta_{s}})-\beta|^{2}\quad(e^{i\theta_{s}}\in B)$:\\
Substituting the singular points into $|\alpha-\sqrt{2}e^{2\pi i\phi}\tilde{f}(e^{i\theta})-\beta|^{2}$, we obtain
\begin{eqnarray}|\alpha-\sqrt{2}e^{2\pi i\phi}\tilde{f}(e^{i\theta})-\beta|^{2}
=\left\{ \begin{array}{ll}
|\beta+i\alpha|^{2}&(\theta=\theta_{1}^{(\pm)}),\\
|\beta-i\alpha|^{2}&(\theta=\theta_{2}^{(\pm)}).
\end{array} \right.\label{(24)}\end{eqnarray} 

\end{enumerate}
Combining the computed items (Lemma $8$, Eqs.(\ref{(23)}), (\ref{(25)}) and (\ref{(24)})) with Eq.(\ref{(21)}), 
we arrive at $|x|\geq1$ case for Theorem \ref{main-result2}.\\

\par\indent
\par\noindent
{\bf Acknowledgments.}
NK acknowledges financial support of the Grant-in-Aid for Scientific
Research (C) of Japan Society for the Promotion of Science (Grant No. $21540116$).
\par\indent
\par\noindent
\nonumsection{References}

\appendix\\
In Appendix A, we give the proof of Theorem $1$.
Using Lemma \ref{wojcik-lemma1}, we compute the generating function of $\Psi_{n} ^{(L)} (0)$.
First, we put $x_n = r^{\ast}_{2n-1}$ and $u=(\omega(-1+\eta i)/2$. Then we have
\begin{align}
\sum_{n=1}^{\infty} \Psi_{2n} ^{(L)} (0) z^{2n} 
&= \frac{1}{\sqrt{2}} \> 
\sum_{n=1}^{\infty} \Biggl\{ \sum_{k=1}^n \sum_{\scriptstyle (a_1, \ldots , a_k) \in (\mathbb{Z}_{>})^k : \atop \scriptstyle a_1 + \cdots + a_k =n} \left( \prod_{j=1}^k x_{a_j} \right) u^k \Biggr\} z^{2n}
\label{(prf1.1)}\\
&= \frac{1}{\sqrt{2}} \> 
\sum_{k=1}^{\infty} \Biggl\{ \sum_{n=k}^{\infty} \sum_{\scriptstyle (a_1, \ldots , a_k) \in (\mathbb{Z}_{>})^k : \atop \scriptstyle a_1 + \cdots + a_k =n} \left( \prod_{j=1}^k x_{a_j} \right)  z^{2n} \Biggr\} u^k  
\nonumber\\
&= \frac{1}{\sqrt{2}} \> 
\sum_{k=1}^{\infty} \left\{ (-1 - z^2 + \sqrt{1 + z^4}) \> u \right\}^k 
\nonumber\\
&= \frac{1}{\sqrt{2}} \> \frac{ (-1 - z^2 + \sqrt{1 + z^4}) u}{1 - (-1 - z^2 + \sqrt{1 + z^4}) u}.\label{prf1.2}
\end{align}
\noindent
Noting the initial coin state $\Psi_{0} ^{(L)} (0) =1/\sqrt{2}$, we obtain 
\begin{align*}
\sum_{n=0}^{\infty} \Psi_{n} ^{L} (0) z^{n} = \frac{1}{\sqrt{2}} \> \frac{1}{1 - Z u},
\end{align*}
where $Z = -1 - z^2 + \sqrt{1 + z^4}$.
In a similar way, we have
\begin{align}
\sum_{n=0}^{\infty} \Psi_{n} ^{R} (0) z^{n} = \frac{\eta i}{\sqrt{2}} \frac{1}{1 - Z u}.\label{(107)}
\end{align}
Equation (\ref{(107)}) leads to the following.
\begin{align*}
\sum_{n=0}^{\infty} \Psi_n ^{(L,\Re)} (0) z^n 
&= \eta \> \sum_{n=0}^{\infty} \Psi_n ^{(R,\Im)} (0) z^n 
= \frac{2+ (C+ \eta S) Z }{\sqrt{2} \{ 2+ 2 (C+ \eta S) Z + Z^2 \}}, 
\\
\sum_{n=0}^{\infty} \Psi_n ^{(L,\Im)} (0) z^n 
&= (-\eta) \> \sum_{n=0}^{\infty} \Psi_n ^{(R,\Re)} (0) z^n 
= \frac{(\eta C-S) Z}{\sqrt{2} \{2+ 2 (C+ \eta S) Z + Z^2 \}}.\label{()}
\end{align*}
We should note that $\Psi_n ^{(A,\Re)} (0)$  ($\Psi_n ^{(A,\Im)} (0)$) is the real(imaginary) part of 
$\Psi_n ^{(A)} (0) \>\> (A= L, R)$, and $C= \cos (2 \pi \phi), \> S= \sin (2 \pi \phi)$. 
Hence, noting $E= C + \eta S$, we get
\begin{align*}
&\sum_{n=0}^{\infty} \Psi_{2n} ^{(L,\Re)} (0) w^n 
= \eta \> \sum_{n=0}^{\infty} \Psi_{2n} ^{(R,\Im)} (0) w^n \\
& \qquad = \frac{4 - 3E + 2(1-E)^2 w + (2 - E) w^2 + (2-E) (1 + w) \sqrt{1+ w^2}}{2 \sqrt{2} \> \left\{ 3 - 2E + 2(1-E)^2 w + (3 - 2E) w^2 \right\}}, \\
&\sum_{n=0}^{\infty} \Psi_{2n} ^{(L,\Im)} (0) w^n 
= (-\eta) \> \sum_{n=0}^{\infty} \Psi_{2n} ^{(R,\Re)} (0) w^n \\
& \qquad = \frac{(S - \eta C) (1 + 2(1-E) w + w^2 - (1 - w) \sqrt{1+ w^2})}{2 \sqrt{2} \> \left\{ 3 - 2E + 2(1-E)^2 w + (3 - 2E) w^2 \right\}}.
\end{align*}
\par\indent
\par\noindent
\begin{lemma}
\label{prf1.1}
Let  $\varphi = \varphi (\eta) = {}^T [1/\sqrt{2},\eta i/\sqrt{2}] \> (\eta =1,-1)$ be the initial coin state. Then we have
\begin{align*}
&\Psi_{2n} ^{(L,\Re)} (0)  
= \eta \Psi_{2n} ^{(R,\Im)} (0)  
\\
& \quad \sim \frac{\sqrt{2} (1 - E)}{3 - 2 E} \> \cos (n \theta_0) 
\times \left\{ I_{(1/4,1)} (\phi) \> I_{\{1\}}(\eta) + I_{(0,3/4)} (\phi) \> I_{\{-1\}}(\eta) \right\},
\\
&\Psi_{2n} ^{(L,\Im)} (0) 
= (- \eta) \Psi_{2n} ^{(R,\Re)} (0) 
\\
& \quad \sim \frac{\sqrt{2} (1 - E)}{3 - 2 E} \frac{S - \eta C}{|S - \eta C|} \> \sin (n \theta_0) 
\times \left\{ I_{(1/4,1)} (\phi) \> I_{\{1\}}(\eta) + I_{(0,3/4)} (\phi) \> I_{\{-1\}}(\eta) \right\},
\end{align*}
where
\begin{align*}
\cos \theta_0 = - \frac{2(1-E)^2}{3-2E}, \quad \sin \theta_0 = \frac{(2-E)|S-C|}{3-2E}.
\end{align*}
Here, $f(n) \sim g(n)$ means $f(n)/g(n)\to1\;(n\to\infty)$.

\end{lemma}
\par\indent

Then the defiition of $r_{2n} (0)$ gives
\begin{align*}
r_{2n} (0) = |\Psi_{2n} ^{(L,\Re)} (0)|^2 + |\Psi_{2n} ^{(L,\Im)} (0)|^2 + |\Psi_{2n} ^{(R,\Re)} (0)|^2 + |\Psi_{2n} ^{(R,\Im)} (0)|^2,
\end{align*}
and we obtain the desired conclusion.

From now on, we explain the outline of the computation for the case of $\eta = 1$.
We can compute for $\eta = -1$ case in a similar way.
Let
\begin{align*}
\sum_{n=0}^{\infty} \Psi_{2n} ^{(L,\Re)} (0) w^n = A_1 (w) + A_2 (w),
\end{align*}
where
\begin{align*}
A_1 (w) 
&= \frac{4 - 3E + 2(1-E)^2 w + (2 - E) w^2}{2 \sqrt{2} \> \left\{ 3 - 2E + 2(1-E)^2 w + (3 - 2E) w^2 \right\}},
\\
A_2 (w) 
&= \frac{(2-E) (1 + w) \sqrt{1+ w^2}}{2 \sqrt{2} \> \left\{ 3 - 2E + 2(1-E)^2 w + (3 - 2E) w^2 \right\}}.
\end{align*}
Next we put the roots  satisfying the quadratic equation:
\begin{align*}
1+\frac{2(1-E)^2}{3-2E}w+w^2=0,
\end{align*}
as $\gamma=e^{i \theta_0}$ and $\overline{\gamma}=e^{-i \theta_0}$.
Then, we see 
\begin{align*}
\cos \theta_0=-\frac{(1-E)^2}{3-2E} \quad (<0).
\end{align*}
Hence we have
\begin{align*}
A_1 (w) = \frac{4 - 3E + 2(1-E)^2 w + (2 - E) w^2}{2 \sqrt{2} \> (3-2E) (w - \gamma) (w - \overline{\gamma})}.
\end{align*}
From now on, we denote $[z^n](f(z))=f_n$, when $f(z)$ has the infinite geometric series
\begin{align*}
f(z) = \sum_{n=0}^{\infty} f_n z^n.
\end{align*}
\par\indent
\par\noindent
Then noting
\begin{align*}
\frac{1}{w -\gamma} &= - \frac{1}{\gamma(1-w/\gamma)} = - \frac{1}{\gamma} \left(\frac{w}{\gamma}\right)^n,
\\
\frac{1}{w -\overline{\gamma}} &= - \frac{1}{\overline{\gamma}(1-w/\overline{\gamma})} = - \frac{1}{\overline{\gamma}} \left(\frac{w}{\overline{\gamma}}\right)^n, 
\end{align*}
we obtain
\begin{align*}
[w^n] \left( \frac{1}{w -\gamma} \right) &= - \gamma^{-(n+1)} = - e^{-i(n+1) \theta_0},
\\
[w^n] \left( \frac{1}{w -\overline{\gamma}} \right) &= - \overline{\gamma}^{-(n+1)} = - \gamma^{n+1} = - e^{i(n+1) \theta_0}.
\end{align*}
Therefore we get
\begin{eqnarray}
[w^n] \left( A_1 (w) \right) 
\!\!\!&\sim 
[w^n] \left( \dfrac{4 - 3E + 2(1-E)^2 \gamma + (2 - E) \gamma^2}{2 \sqrt{2} \> (3-2E)  (\gamma - \overline{\gamma}) (w - \gamma)} \right) 
\nonumber\\
\qquad\qquad&+ [w^n] \left( \dfrac{4 - 3E + 2(1-E)^2 \overline{\gamma} + (2 - E) \overline{\gamma}^2}{2 \sqrt{2} \> (3-2E)  (\overline{\gamma} - \gamma) (w - \overline{\gamma})} \right).
\label{jobs1}
\end{eqnarray}
The above discussion comes from \cite{flajolet}, for example.
Here Eq. (\ref{jobs1}) gives
\begin{align*}
[w^n] \left( A_1 (w) \right) 
&\sim \frac{1-E}{2 \sqrt{2} \> (3-2E)} \times 
[w^n] \left( \frac{1- \gamma^2}{(\gamma - \overline{\gamma}) (w - \gamma)} + 
\frac{1 - \overline{\gamma}^2}{(\overline{\gamma} - \gamma) (w - \overline{\gamma})} \right)
\\
&= - \frac{1-E}{2 \sqrt{2} \> (3-2E)} \times \left\{ \frac{1- \gamma^2}{\gamma - \overline{\gamma}} \> \gamma^{-(n+1)} + \frac{1- \overline{\gamma}^2}{\overline{\gamma} - \gamma} \> \overline{\gamma}^{-(n+1)}\right\}
\\
&= - \frac{1-E}{\sqrt{2} \> (3-2E)} \times \Re \left( \frac{1- \gamma^2}{\gamma - \overline{\gamma}} \> \gamma^{-(n+1)} \right)
\\
&= - \frac{1-E}{\sqrt{2} \> (3-2E)} \times \Re \left( \frac{\overline{\gamma} - \gamma}{\gamma - \overline{\gamma}} \> \gamma^{-n} \right)
\\
&= \frac{1-E}{2 \sqrt{2} \> (3-2E)} \> \cos (n \theta_0).\end{align*}
In a similar fashion, noting
\begin{align*}
\sqrt{1+\gamma^2}
= \sqrt{-2 \cos \theta_0} \left( \sin \left( \frac{\theta_0}{2} \right) - i \cos \left( \frac{\theta_0}{2} \right)\right), \quad 
\sqrt{1+\overline{\gamma}^2}
= \overline{\sqrt{1+\gamma^2}},
\end{align*}
we have
\begin{align}
[w^n] \left( A_2 (w) \right) 
&\sim - \frac{2-E}{2 \sqrt{2} \> (3-2E)} \times \left\{ \frac{(1+ \gamma) \sqrt{1+\gamma^2}}{\gamma - \overline{\gamma}} \> \gamma^{-(n+1)} 
\right.
\nonumber\\
& \qquad \qquad \qquad \qquad \qquad \qquad \left.
+ \frac{(1+ \overline{\gamma}) \sqrt{1+\overline{\gamma}^2}}{\overline{\gamma} - \gamma} \> \overline{\gamma}^{-(n+1)} \right\}
\nonumber\\
&= - \frac{2-E}{\sqrt{2} \> (3-2E)} \times \Re \left( \frac{(1+ \gamma) \sqrt{1+\gamma^2}}{\gamma - \overline{\gamma}} \> \gamma^{-(n+1)} \right)
\nonumber\\
&= - \frac{(2-E)\sqrt{-2 \cos \theta_0}}{\sqrt{2} \> (3-2E)} \times \Re \left( \frac{(\overline{\gamma} +1) (-i) e^{i \theta_0/2}}{\gamma - \overline{\gamma}} \> \gamma^{-n} \right)
\\
&= \frac{(2-E)\sqrt{-\cos \theta_0}}{3-2E} \times \left( \frac{\cos (n \theta_0)}{2 \sin(\theta_0/2)} \right).
\label{(108)}\end{align}
Moreover, substituting
\begin{align*}
\sqrt{-\cos \theta_0} = \frac{|1-E|}{\sqrt{3-2E}}, \quad \sin \left( \frac{\theta_0}{2} \right) = \frac{2-E}{\sqrt{2} \sqrt{3-2E}}
\end{align*}
into Eq.(\ref{(108)}), we see
\begin{align*}
[w^n] \left( A_2 (w) \right) \sim \frac{|1-E|}{\sqrt{2}(3-2E)} \cos (n \theta_0).\end{align*}
Thus, we obtain
\begin{align*}
\Psi_{2n} ^{(L,\Re)} (0) 
&= [w^n] \left( A_1 (w) + A_2 (w) \right)
\\
& \sim \frac{1-E + |1-E|}{\sqrt{2}(3-2E)} \cos (n \theta_0)= \frac{\sqrt{2}(1-E)}{3-2E} \cos (n \theta_0) I_{(1/4,1)} (\phi).
\end{align*}
Here we should note that the condition $1-E>0$ is equivalent to $\phi \in (1/4,1)$.
Moreover, 
\begin{align*}\Psi_{2n} ^{(L,\Re)} (0) = \Psi_{2n} ^{(R,\Im)} (0).\end{align*} 
holds.
Now we put
\begin{align*}
\sum_{n=0}^{\infty} \Psi_{2n} ^{(L,\Im)} (0) w^n = B_1 (w) + B_2 (w),\end{align*}
where
\begin{align*}
B_1 (w) 
&= \frac{(S-C)(1 + 2(1-E) w + w^2)}{2 \sqrt{2} \> \left\{ 3 - 2E + 2(1-E)^2 w + (3 - 2E) w^2 \right\}},
\\
B_2 (w) 
&= - \frac{(S-C) (1 - w) \sqrt{1+ w^2}}{2 \sqrt{2} \> \left\{ 3 - 2E + 2(1-E)^2 w + (3 - 2E) w^2 \right\}}.
\end{align*}
First, we consider $B_1 (w)$.
Noting \begin{align*}
\cos \theta_0 = - \frac{(1-E)^2}{3-2E}, \quad \sin \theta_0 = \frac{|S-C| (2-E)}{3-2E},
\end{align*}
we have
\begin{align*}
[w^n] \left( B_1 (w) \right) 
&\sim \frac{C-S}{\sqrt{2} \> (3-2E)} \times 
\Re \left( \frac{\gamma + \overline{\gamma} +2(1-E)}{\gamma - \overline{\gamma}} \> \gamma^{-n} \right)
\\
&\sim \frac{S-C}{\sqrt{2} \> (3-2E)} \times 
\frac{\cos \theta_0 + 1-E}{\sin \theta_0} \times \sin (n \theta_0)
\\
&= \frac{1-E}{\sqrt{2} \> (3-2E)} \times \frac{S-C}{|S-C|} \times \sin (n \theta_0). 
\end{align*}
In a similar way, noting
\begin{align*}
\sqrt{1+\gamma^2}
= \sqrt{-2 \cos \theta_0} \left( \sin \left( \frac{\theta_0}{2} \right) - i \cos \left( \frac{\theta_0}{2} \right)\right), \quad 
\sqrt{1+\overline{\gamma}^2}
= \overline{\sqrt{1+\gamma^2}},
\end{align*}
we get
\begin{align}
[w^n] \left( B_2 (w) \right) 
&\sim \frac{S-C}{\sqrt{2} \> (3-2E)} \Re \left( \frac{(\overline{\gamma} - 1) \sqrt{-2 \cos \theta_0} ( \sin (\theta_0/2) - i \cos (\theta_0/2))}{\gamma - \overline{\gamma}}  \> \gamma^{-n} \right)
\nonumber\\
&= \frac{S-C}{2 \> (3-2E)} \times \frac{\sqrt{-\cos \theta_0}}{\cos (\theta_0/2)} \times \sin (n \theta_0).\label{(109)}
\end{align}
Accordingly, substituting
\begin{align*}
\sqrt{-\cos \theta_0} = \frac{|1-E|}{\sqrt{3-2E}}, \quad \cos \left( \frac{\theta_0}{2} \right) = \frac{|S-C|}{\sqrt{2} \sqrt{3-2E}}
\end{align*}
into Eq. (\ref{(109)}), we see
\begin{align*}
[w^n] \left( B_2 (w) \right) \sim \frac{|1-E|}{\sqrt{2}(3-2E)} \times \frac{S-C}{|S-C|} \times \sin (n \theta_0).
\end{align*}
Hence, we have
\begin{align*}
\Psi_{2n} ^{(L,\Im)} (0) 
&= [w^n] \left( B_1 (w) + B_2 (w) \right)
\\
& \sim \frac{1-E + |1-E|}{\sqrt{2}(3-2E)} \> \frac{S-C}{|S-C|} \> \sin (n \theta_0) 
\\
&= \frac{\sqrt{2}(1-E)}{3-2E} \> \frac{S-C}{|S-C|} \> \sin (n \theta_0)  I_{(1/4,1)} (\phi).
\end{align*}
Here we should note that the condition $1-E>0$ is equivalent to $\phi \in (1/4,1)$.
Further, $\Psi_{2n} ^{(L,\Im)} (0) = - \Psi_{2n} ^{(R,\Re)} (0)$ holds.
Therefore, we obtain
\par\indent
\par\noindent
\begin{lemma}
\label{prf1.2}
For $\varphi = {}^T [1/\sqrt{2}, i/\sqrt{2}]$, we have
\begin{align*}
\Psi_{2n} ^{(L,\Re)} (0) 
&= \Psi_{2n} ^{(R,\Im)} (0) \sim \frac{\sqrt{2}(1-E_{+})}{3-2E_{+}} \cos (n \theta_0) I_{(1/4,1)} (\phi),
\\
\Psi_{2n} ^{(L,\Im)} (0) 
&= - \Psi_{2n} ^{(R,\Re)} (0) \sim \frac{\sqrt{2}(1-E_{+})}{3-2E_{+}} \> \frac{S-C}{|S-C|} \> \sin (n \theta_0)  I_{(1/4,1)} (\phi).
\end{align*}
where $E_{+}=C+S= \cos (2 \pi \phi) +  \sin (2 \pi \phi)$.
\end{lemma}
\par\indent
\par\noindent
Thus
\begin{align*}
r_{2n} (0) = |\Psi_{2n} ^{(L,\Re)} (0)|^2 + |\Psi_{2n} ^{(L,\Im)} (0)|^2 + |\Psi_{2n} ^{(R,\Re)} (0)|^2 + 
|\Psi_{2n} ^{(R,\Im)} (0)|^2
\end{align*}
gives
\begin{align*}
\lim_{n \to \infty} r_{2n} (0) 
&= \lim_{n \to \infty} 2 \times \frac{2(1-E_{+})^2}{(3-2E_{+})^2} \> \left( \cos^2 (n \theta_0) + \sin^2 (n \theta_0) \right) \> I_{(1/4,1)} (\phi)
\\
&= \frac{4(1-E_{+})^2}{(3-2E_{+})^2} \> I_{(1/4,1)} (\phi).
\end{align*}
Noting the definition of
\begin{align*}\mu_{\infty}(0)=\lim_{n\to\infty}r_{2n}(0)/2,\label{()}\end{align*}
we obtain 
\par\indent
\par\noindent
\begin{lemma}
\label{prf1.3}
For $\varphi = {}^T [1/\sqrt{2}, i/\sqrt{2}]$, we have
\begin{align*}
\mu_{\infty} (0) = \frac{2(1-E_{+})^2}{(3-2E)^2} \> I_{(1/4,1)} (\phi) = 2 \times 
\left( \frac{1 - \sqrt{2}C_{-}}{3 - 2 \sqrt{2}C_{-}} \right)^2 \> I_{(1/4,1)} (\phi).
\end{align*}
where
\begin{align*}
E_{+} &= C+S= \cos (2 \pi \phi) +  \sin (2 \pi \phi), 
\\
C_{-} 
&= \cos \left( 2 \pi \phi - \frac{\pi}{4} \right) = \frac{\sqrt{2}}{2} \left\{ \cos ( 2 \pi \phi) + \sin ( 2 \pi \phi) \right\}.
\end{align*}
\end{lemma}
Next, similarly we show the results for the case of $\eta = -1$, that is, $\varphi = {}^T [1/\sqrt{2}, -i/\sqrt{2}]$ case.
\par\indent
\par\noindent
\begin{lemma}
\label{prf1.4}
For $\varphi = {}^T [1/\sqrt{2}, -i/\sqrt{2}]$, we have
\begin{align*}
\Psi_{2n} ^{(L,\Re)} (0) 
&= - \Psi_{2n} ^{(R,\Im)} (0) \sim \frac{\sqrt{2}(1-E_{-})}{3-2E_{-}} \cos (n \theta_0) I_{(0,3/4)} (\phi),
\\
\Psi_{2n} ^{(L,\Im)} (0) 
&= \Psi_{2n} ^{(R,\Re)} (0) \sim \frac{\sqrt{2}(1-E_{-})}{3-2E_{-}} \> \frac{S+C}{|S+C|} \> \sin (n \theta_0)  I_{(0,3/4)} (\phi),
\end{align*}
Moreover,
\begin{align*}
\mu_{\infty} (0) = \frac{2(1-E_{-})^2}{(3-2_{-}E)^2} \> I_{(0,3/4)} (\phi) = 2 \times \left( \frac{1 - \sqrt{2}C_{+}}
{3 - 2 \sqrt{2}C_{+}} \right)^2 \> I_{(0,3/4)} (\phi),
\end{align*}
where
\begin{align*}
E_{-} 
&= C-S= \cos (2 \pi \phi) -  \sin (2 \pi \phi), 
\\
C_{-} 
&= \cos \left( 2 \pi \phi - \frac{\pi}{4} \right) = \frac{\sqrt{2}}{2} \left\{ \cos ( 2 \pi \phi) + \sin ( 2 \pi \phi) \right\}.
\end{align*}
\end{lemma}
\par\indent
\par\noindent
Therefore, 
for $\varphi = \varphi (\eta) = {}^T [1/\sqrt{2},\eta i/\sqrt{2}] \> \;(\eta =1,-1)$, we have
\begin{align*}
&\Psi_{2n} ^{(L,\Re)} (0)  
= \eta \Psi_{2n} ^{(R,\Im)} (0)  
\\
& \quad \sim \frac{\sqrt{2} (1 - E)}{3 - 2 E} \> \cos (n \theta_0) 
\times \left\{ I_{(1/4,1)} (\phi) \> I_{\{1\}}(\eta) + I_{(0,3/4)} (\phi) \> I_{\{-1\}}(\eta) \right\},
\\
&\Psi_{2n} ^{(L,\Im)} (0) 
= (- \eta) \Psi_{2n} ^{(R,\Re)} (0) 
\\
& \quad \sim \frac{\sqrt{2} (1 - E)}{3 - 2 E} \frac{S - \eta C}{|S - \eta C|} \> \sin (n \theta_0) 
\times \left\{ I_{(1/4,1)} (\phi) \> I_{\{1\}}(\eta) + I_{(0,3/4)} (\phi) \> I_{\{-1\}}(\eta) \right\},
\end{align*}
where
\begin{align*}
\cos \theta_0 = - \frac{2(1-E)^2}{3-2E}, \quad \sin \theta_0 = \frac{(2-E)|S-C|}{3-2E}.
\end{align*}
Thus, we obtain
\begin{eqnarray}
\mu_{\infty} (0) 
\!\!\!&=&\!\!\!\lim_{n \to \infty} \> \frac{r_{2n} (0)}{2}
\nonumber\\
\!\!\!&=&\!\!\!2 \left( \frac{1- \sqrt{2}C_{-}}{3- 2\sqrt{2}C_{-}} \right)^2  I_{(1/4,1)} (\phi) \> I_{\{1\}}(\eta)
+ 2 \left( \frac{1- \sqrt{2}C_{+}}{3- 2\sqrt{2}C_{+}} \right)^2 I_{(0,3/4)} (\phi) \> I_{\{-1\}}(\eta).\label{(110)}
\end{eqnarray}
Therefore, the proof is complete.

\appendix\\
In Appendix B, we present the proof of Lemma $7$ which gives the singular points of $\tilde{\Xi}_{x}(z)$ for the Wojcik model.
We should recall that the singular points of $\tilde{\Xi}_{x}(z)$ come from $1/\tilde{\Lambda}_{0}(z)$ part.
Note 
\[\tilde{\Lambda}_{0}(e^{i\theta})=1-\sqrt{2}\omega\tilde{f}(e^{i\theta})+\omega^{2}
\{\tilde{f}(e^{i\theta})\}^{2},\]
where $\;\tilde{f}(e^{i\theta})=e^{i(\theta+\tilde{\phi}(\theta))}$ with
\[\left\{
\begin{array}{l}
\sin\tilde{\phi}(\theta)=\operatorname{sgn}(\sin\theta)\sqrt{2\sin\theta^{2}-1},\\
\cos\tilde{\phi}(\theta)=\sqrt{2}\cos\theta,
\end{array}
\right.\]
and $\omega=e^{2\pi i\phi}\;(\phi\in(0,1))$. 
Here we need to derive all $\theta$ satisfying
\begin{align}1-\sqrt{2}\omega\tilde{f}(e^{i\theta})+\omega^{2}\{\tilde{f}
(e^{i\theta})\}^{2}=0.\label{(28)}\end{align}
Equation (\ref{(28)}) implies that we have $\tilde{f}(e^{i\theta})=e^{-2\pi i \phi+\pi/4}$ or 
$\tilde{f}(e^{i\theta})=e^{-2\pi i \phi-\pi/4}$.
Noting $\tilde{f}(e^{i\theta})=e^{i(\theta+\tilde{\phi}(\theta))}$, we have the two cases as follows:
\par\indent
\par\noindent
\begin{enumerate}
\item ${e^{i(\theta+\tilde{\phi}(\theta))}}=e^{-2\pi i \phi-\pi/4}$ case.\\
In this case, we have 
\begin{align*}
e^{i(2\pi\phi+\theta+\tilde{\phi}(\theta))}=e^{-\pi i/4}.
\end{align*}
Hence we see
\begin{align}
2\pi\phi+\theta+\tilde{\phi}(\theta)=-\dfrac{\pi}{4}.\label{(30)}
\end{align}
Noting $\sqrt{2}\cos\theta=\cos\tilde{\phi}(\theta)$, Eq.(\ref{(30)}) gives
\begin{eqnarray}
\sqrt{2}\cos\theta=\cos\tilde{\phi}(\theta)=\cos\left(-\dfrac{\pi}{4}-2\pi \phi-\theta\right)=\cos\left(\dfrac{\pi}{4}+2\pi\phi+\theta\right).\label{(31)}
\end{eqnarray}
Putting $\epsilon=2\pi\phi+\pi/4$, Eq. (\ref{(31)}) becomes
\begin{align*}
\sqrt{2}\cos\theta=\cos\epsilon\cos\theta-\sin\epsilon\sin\theta.
\end{align*}
Therefore we have
\begin{align*}
\cos\theta=\dfrac{\sin\epsilon}{\cos\epsilon-\sqrt{2}}\sin\theta.
\end{align*}
Noting $\sin^{2}\theta+\cos^{2}\theta=1$, we get
\begin{align*}
\sin\theta=\pm\dfrac{\cos\epsilon}{\sqrt{3-2\sqrt{2}\cos\epsilon}}.
\end{align*}
Then, we obtain
\begin{align*}
\left\{\begin{array}{l}
(\cos\theta^{(1)},\sin\theta^{(1)})=\left(\dfrac{\sin\epsilon}{\sqrt{3-2\sqrt{2}\cos\epsilon}},\dfrac{\cos\epsilon-\sqrt{2}}{\sqrt{3-2\sqrt{2}\cos\epsilon}}\right),\\
(\cos\theta^{(2)},\sin\theta^{(2)})=\left(-\dfrac{\sin\epsilon}{\sqrt{3-2\sqrt{2}\cos\epsilon}},-\dfrac{\cos\epsilon-\sqrt{2}}{\sqrt{3-2\sqrt{2}\cos\epsilon}}\right).
\end{array} \right.
\end{align*} 

\item ${e^{i(\theta+\tilde{\phi}(\theta))}}=e^{-2\pi i \phi+\pi/4}$ case.\\
In this case,  
\begin{align*}
e^{i(2\pi\phi+\theta+\tilde{\phi}(\theta))}=e^{\pi i/4}
\end{align*}
holds.
Hence we have
\begin{align}
2\pi\phi+\theta+\tilde{\phi}(\theta)=\dfrac{\pi}{4}.\label{(37)}
\end{align}
Noting $\sqrt{2}\cos\theta=\cos\tilde{\phi}(\theta)$, Eq.(\ref{(37)}) yields
\begin{eqnarray}
\sqrt{2}\cos\theta=\cos\tilde{\phi}(\theta)=\cos\left(\dfrac{\pi}{4}-2\pi \phi-\theta\right)=\cos\left(2\pi\phi-\dfrac{\pi}{4}+\theta\right).\label{(38)}
\end{eqnarray}
Putting $\tilde{\epsilon}=2\pi\phi-\pi/4$, Eq. (\ref{(38)}) becomes
\begin{align*}
\sqrt{2}\cos\theta=\cos\tilde{\epsilon}\cos\theta-\sin\tilde{\epsilon}\sin\theta.
\end{align*}
Thus, we have
\begin{align*}
\cos\theta=\dfrac{\sin\tilde{\epsilon}}{\cos\tilde{\epsilon}-\sqrt{2}}\sin\theta.
\end{align*}
Noting $\sin^{2}\theta+\cos^{2}\theta=1$, we get
\begin{align*}
\sin\theta=\pm\dfrac{\cos\tilde{\epsilon}-\sqrt{2}}{\sqrt{3-2\sqrt{2}\cos\tilde{\epsilon}}}.
\end{align*}
Therefore, we obtain
\begin{align*}
\left\{\begin{array}{l}
(\cos\theta^{(3)},\sin\theta^{(3)})=\left(\dfrac{\sin\tilde{\epsilon}}{\sqrt{3-2\sqrt{2}\cos\tilde{\epsilon}}},\dfrac{\cos\tilde{\epsilon}-\sqrt{2}}{\sqrt{3-2\sqrt{2}\cos\tilde{\epsilon}}}\right),\\
(\cos\theta^{(4)},\sin\theta^{(4)})=\left(-\dfrac{\sin\tilde{\epsilon}}{\sqrt{3-2\sqrt{2}\cos\tilde{\epsilon}}},-\dfrac{\cos\tilde{\epsilon}-\sqrt{2}}{\sqrt{3-2\sqrt{2}\cos\tilde{\epsilon}}}\right).
\end{array} \right.
\end{align*} 
\end{enumerate}
\end{document}